\documentclass[aps, prb, acs, amsmath, twocolumn, longbibliography, superscriptaddress, footinbib, 10pt, final]{revtex4-2}

\pdfoutput=1 
\usepackage[utf8]{inputenc}
\usepackage{tabularx}
\usepackage{bm}
\usepackage{gensymb}
\usepackage{ifthen}
\usepackage{float}
\usepackage{mathtools}
\usepackage{booktabs}
\usepackage[version=4]{mhchem} 
\usepackage{physics}
\usepackage{amssymb, amsmath}
\usepackage{hyphenat}
\usepackage{comment}
\usepackage{graphicx} 
\usepackage{epstopdf}  
\usepackage{blindtext}
\usepackage{siunitx} 
\usepackage[final]{microtype} 
\usepackage{setspace}
\usepackage[T1]{fontenc}
\usepackage{lmodern}
\usepackage{xcolor}
\usepackage[colorlinks=true,linkcolor=black, citecolor=blue, urlcolor=blue]{hyperref}
\usepackage[utf8]{inputenc}
\usepackage[acronym]{glossaries}
\usepackage[l2tabu, orthodox]{nag}
\usepackage{makecell}
\usepackage{color,soul}
\usepackage{adjustbox}
\usepackage{environ} 
\newcommand{\kp}{$\bm{k \cdot p}$}  
\newcommand{\GeSn}[2]{Ge$_{#1}$Sn$_{#2}$}

\begin{document}

\def\figureautorefname{Fig.} 
\def\tableautorefname{Table} 
\newcommand{\fref}[2]{\autoref{#1}\textcolor{blue}{#2}}

\newcommand{\paperSection}[2][normal]{
    \ifthenelse{\equal{#1}{normal}}{
        \medskip\noindent{\textbf{#2}}\newline
    }{
        \noindent\textbf{#2}\newline
    }
}

 
\newcommand{\RomanNumeralCaps}[1]
    {\MakeUppercase{\romannumeral #1}}

\newcommand{\polydept}{Department of Engineering Physics, \'Ecole Polytechnique de Montr\'eal, C.P. 6079, Succ. Centre-Ville, Montr\'eal, Qu\'ebec, Canada H3C 3A7}

\newcommand{\citechange}[1]{\textcolor{red}{#1}}
\newcommand{\SiGeSn}[3]{Si$_{#1}$Ge$_{#2}$Sn$_{#3}$}
\newcommand{\SipGeSn}[3]{(Si)$_{#1}$Ge$_{#2}$Sn$_{#3}$}
\newcommand{\SiGe}[2]{Si$_{#1}$Ge$_{#2}$}
\newcommand{\quot}[1]{``#1''} 
\newcommand{\ac}[1]{\gls*{#1}}
\newcommand{\acc}[1]{\Gls*{#1}}
\newcommand{\RNum}[1]{\uppercase\expandafter{\romannumeral #1\relax}}

\newcommand{\citesupp}[1]{
    \footnote{See Supplemental Material at \textbf{[URL will be inserted by publisher]} for \textcolor{red}{details of the experimental setup}, which includes Refs. #1}
}


\newcommand{\figCiteChange}[1]{\colorbox{yellow!50!white}{#1}}
\DeclareSIUnit{\million}{\text{million}}


\newcommand\vertarrowbox[3][3ex]{%
  \begin{array}[t]{@{}c@{}} #2 \vspace{1ex}\\
  \left\uparrow\vcenter{\hrule height #1}\right.\kern-\nulldelimiterspace\\
  \makebox[0pt]{#3}
  \end{array}%
}

\NewEnviron{myequation}[1]{%
\begin{equation}
\scalebox{#1}{$\BODY$}
\end{equation}
}

\newcommand{\suppfref}[2]{\autoref{#1}\textcolor{blue}{#2}}

\newcommand\suppFigRef[3]{%
    \ifthenelse{\equal{\includeSI}{1}}{%
        \let\tempfigureautorefname\figureautorefname%
        \renewcommand\figureautorefname{Fig.}%
        \suppfref{#1}{#2}%
        \let\figureautorefname\tempfigureautorefname%
    }{\textcolor{blue}{Fig. S#3}}%
}

\newcommand\suppFigsRef[3]{%
    \ifthenelse{\equal{\includeSI}{1}}{%
        \let\tempfigureautorefname\figureautorefname%
        \renewcommand\figureautorefname{Figs.}%
        \suppfref{#1}{#2}%
        \let\figureautorefname\tempfigureautorefname%
    }{\textcolor{blue}{Figs. S#3}}%
}

\newcommand\suppTabRef[3]{%
    \ifthenelse{\equal{\includeSI}{1}}{%
        \let\temptableautorefname\tableautorefname%
        \renewcommand\tableautorefname{Table}%
        \suppfref{#1}{#2}%
        \let\tableautorefname\temptableautorefname%
    }{\textcolor{blue}{Table S#3}}%
}

\def\blankpage{%
      \clearpage%
      \thispagestyle{empty}%
      \null%
      \clearpage}

\epstopdfsetup{update} 
\epstopdfsetup{outdir=./figures/}
\epstopdfsetup{suffix=-generated}

\newacronym{se}{SE}{spectroscopic ellipsometry}
\newacronym{apt}{APT}{atom probe tomography}
\newacronym{fib}{FIB}{focused ion beam}
\newacronym{hrxrd}{HRXRD}{high\hyp{}resolution xray diffraction}
\newacronym{rsm-xrd}{RSM\hyp{}XRD}{reciprocal space mapping X\hyp{}ray diffraction spectroscopy}
\newacronym{rsm}{RSM}{reciprocal space mapping}
\newacronym{fwhm}{FWHM}{full\hyp{}width half maximum}
\newacronym{edsx}{EDSX}{energy\hyp{}dispersive X\hyp{}ray spectroscopy}
\newacronym{tem}{TEM}{transmission electron microscope}
\newacronym{hrtem}{HRTEM}{High\hyp{}resolution transmission electron microscopy}
\newacronym{stem}{STEM}{scanning transmission electron microscopy}
\newacronym{mocvd}{MOCVD}{metal\hyp{}organic chemical vapor deposition}
\newacronym{cvd}{CVD}{chemical vapor deposition}
\newacronym{cmos}{CMOS}{complementary metal\hyp{}oxide\hyp{}semiconductor}
\newacronym{haadf-stem}{HAADF\hyp{}STEM}{high\hyp{}angle annular dark field scanning transmission electron microscopy}
\newacronym{rms}{RMS}{surface roughness}
\newacronym{eels}{EELS}{electron energy loss spectroscopy}
\newacronym{rta}{RTA}{rapid thermal annealing}
\newacronym{lta}{LTA}{laser thermal annealing}
\newacronym{ia}{IA}{isothermal annealing}
\newacronym{rheed}{RHEED}{reflection high-energy electron diffraction}
\newacronym{peem}{PEEM}{photoemission electron microscopy}
\newacronym{leed}{LEED}{low\hyp{}energy electron diffraction}
\newacronym{sims}{SIMS}{secondary ion mass spectroscopy}

\newacronym{afm}{AFM}{atomic force microscopy}
\newacronym[\glslongpluralkey={misfit dislocations}]{md}{MD}{misfit dislocation}
\newacronym{td}{TD}{threading dislocation}
\newacronym{fib}{FIB}{focused\hyp{}ion beam}
\newacronym{ctlm}{CTLM}{circular transfer length method}
\newacronym{tlm}{TLM}{transfer length method}
\newacronym{rtse}{RTSE}{room temperature spectroscopic ellipsometry}
\newacronym{vs}{VS}{virtual substrate}
\newacronym{pecvd}{PECVD}{plasma enhanced chemical vapor deposition}
\newacronym{aoi}{AOI}{angle of incidence}
\newacronym[\glslongpluralkey={critical\hyp{}points}]{cp}{CP}{critical\hyp{}point}
\newacronym{cppb}{CPPB}{critical point parabolic band}

\title{Mid-Infrared Detectors and Imagers Integrating All-Group IV Nanowires}

\author{Lu Luo} 
\affiliation{\polydept{}}

\author{Mahmoud RM Atalla}  
\affiliation{\polydept{}}

\author{Simone Assali} 
\affiliation{\polydept{}}

\author{Sebastian Koelling} 
\affiliation{\polydept{}}

\author{G\'erard Daligou} 
\affiliation{\polydept{}}

\author{Oussama Moutanabbir}
\affiliation{\polydept{}}

\begin{abstract}
\medskip
Cost-effective mid-wave infrared (MWIR) optoelectronic devices are of utmost importance to a plethora of applications such as night vision, thermal sensing, autonomous vehicles, free-space communication, and spectroscopy. To this end, leveraging the ubiquitous silicon-based processing has emerged as a powerful strategy that can be accomplished through the use of group IV germanium-tin (GeSn) alloys. Indeed, due to their compatibility with silicon and their tunable bandgap energy covering the entire MWIR range, GeSn semiconductors are frontrunner platforms for compact and scalable MWIR technologies. However, the GeSn large lattice parameter has been a major hurdle limiting the quality of GeSn epitaxy on silicon wafers. Herein, it is shown that sub-20 nm Ge nanowires (NWs) provide effective compliant substrates to grow Ge$_{1-x}$Sn$_{x}$ alloys with a composition uniformity over several micrometers with a very limited build-up of the compressive strain. Ge/Ge$_{1-x}$Sn$_{x}$ core/shell NWs with Sn content spanning the 6 to 18 at.$\%$ range are demonstrated and integrated in photoconductive devices exhibiting a high signal-to-noise ratio at room temperature and a tunable cutoff wavelength covering the 2.0 $\mu$m to 3.9 $\mu$m range. Additionally, the processed NW-based detectors were used in uncooled imagers enabling the acquisition of high-quality images under both broadband and laser illuminations without a lock-in technique.
\end{abstract}
\maketitle

\section{INTRODUCTION}\label{sec:intro}
Due to their narrow and tunable direct band, nonequilibrium group IV Ge$_{1-x}$Sn$_{x}$ alloys have attracted considerable interest as versatile silicon-compatible semiconductors for mid-wave infrared (MWIR) photonics and optoelectronics.\cite{moutanabbir2021monolithic,buca2022room,atalla2023extended,chang2022mid,chretien2019gesn,chretien2022room,atalla2021all,atalla2022high,elbaz2020ultra,joo20211d,jung2022optically,li202130,liu2022sn,luo2022extended,marzban2022strain,talamas2021cmos,tran2019si,xu2019high,zhou2020electrically,daligou2023group} Epitaxial methods including molecular beam epitaxy (MBE) and chemical vapor deposition (CVD) are currently broadly adopted to grow these alloys on silicon wafers using predominantly Ge as an interlayer.\cite{moutanabbir2021monolithic} However, these epitaxial Ge$_{1-x}$Sn$_{x}$ thin films still suffer inherent limitations due to the large lattice mismatch between Ge and Sn ($\approx$14.7\%) in addition to the low solubility of Sn in Ge ($\approx$1 at.\%). These limitations are further exacerbated as Ge$_{1-x}$Sn$_{x}$ layers and heterostructures with Sn contents at least one order of magnitude higher than the solubility are needed for device structures relevant to MWIR applications. In this regime, the as-grown layers are typically under a significant compressive strain, which impacts the bandgap directness and increases its energy at the $\Gamma$ point, thus hindering the device performance and limiting the covered range of the MWIR spectrum. This compressive strain build-up not only affects the band structure but also limits the incorporation of Sn atoms in the growing layer, making the control of Sn content a daunting task.\cite{assali2019strain} 

In recent years, several strategies have been pursued to mitigate these challenges. For instance, the growth using compositionally graded Ge$_{1-x}$Sn$_{x}$ buffers, where the amount of Sn is controlled by temperature and precursors flow, was found to be effective in partially relaxing the compressive strain by promoting the nucleation and glide of misfit dislocations in the underlying lower Sn content layers, while enhancing the Sn incorporation and preserving the high-quality of the topmost Sn-rich layer.\cite{aubin2017growth,dou2018investigation,assali2018atomically} Although this growth protocol has been successful in producing device-quality materials, the high density of extended defects in the underlying layers remains a source of non-radiative recombination centers and leakage current in light emitters\cite{buca2022room,chretien2022room,elbaz2020ultra,joo20211d,jung2022optically,marzban2022strain,zhou2020electrically,chang2022mid} and photodetectors.\cite{atalla2022high,atalla2023dark,liu2022sn,luo2022extended,talamas2021cmos,tran2019si,xu2019high}. Methods for defect and strain management using layer transfer, under etching, nanomembrane release, or wafer bonding were subsequently introduced to alleviate the harmful effects of lattice mismatch-induced extended defects leading to a clear improvement in the device performance.\cite{elbaz2020reduced,thai2018gesn,chretien2019gesn,kim2022enhanced,stange2016optically,elbaz2020ultra,burt2021strain,atalla2021all} Notwithstanding this progress, the growth difficulties still impact the nature of material and device structures obtained using these post-growth processing methods. Notably, the layer thickness with high compositional uniformity becomes increasingly small as the Sn content exceeds 10 at.\%, \cite{dou2018investigation,assali2018atomically} thus limiting the potential of Ge$_{1-x}$Sn$_{x}$ for device applications. Nanowires (NWs) provide, in principle, a promising solution to form GeSn alloys with higher crystallinity, primarily due to the inherent enhanced strain relaxation along their free sidewall facets.\cite{galluccio2020field,sistani2018electrical,seifner2019epitaxial} Indeed, the use of Ge NWs as substrates to grow Ge$_{1-x}$Sn$_{x}$ has been explored to form heterostructures with limited defects despite the large lattice mismatch.\cite{meng2016core,assali2017growth} However, the current studies on Ge/Ge$_{1-x}$Sn$_{x}$ core/shell NWs employed a Ge core diameter of 50 nm or larger, systematically resulting in significant compositional fluctuations and defect formation.\cite{meng2016core,assali2019strain,lentz2023local} These hurdles can be mitigated or eliminated by shrinking the Ge core diameter which can enable an enhanced defect-free strain relaxation.\cite{assali2020kinetic,royo2017review} As a matter of fact, modeling studies suggested that a Ge core diameter below 20 nm provides a compliant substrate thereby preventing extended structural defects and Sn segregation,\cite{albani2018critical,assali2020kinetic,kim2023short} which can lead to Ge$_{1-x}$Sn$_{x}$ shells with better content uniformity while minimizing the compressive strain.\cite{kim2023short} \\

Herein, this work puts to test these theoretical predictions and demonstrates the growth of Ge/Ge$_{1-x}$Sn$_{x}$ core/shell NWs using sub-20 nm Ge core leading to a precise control of Sn content uniformity. Indeed, the optimal growth conditions yield NWs with a temperature-controlled Sn composition ranging from 5.6 to 18.0 at.\% with a structural and compositional uniformity along several micrometers in the NW growth direction. This scale of control over the crystalline quality and Sn content is unmatched in thin-film growth. Moreover, the obtained NWs were integrated into MWIR photodetectors and imagers operating at room temperature with Sn content-sensitive cutoff wavelengths in the 2.0-3.9 $\mu$m range. In addition, high-quality images are demonstrated using uncooled NW-based single-pixel imager under broadband and laser illuminations without a lock-in amplifier technique.


\section{Results}\label{sec:methods}
\textbf{Growth and characterization of Ge/Ge$_{1-x}$Sn$_{x}$ core/shell NWs}.
Ge/Ge$_{1-x}$Sn$_{x}$ core/shell NWs were grown on Ge (111) substrates in a chemical vapor deposition (CVD) reactor following a two-step growth protocol. First, Ge core NWs with an average diameter of 20 nm or smaller were grown by following the vapor-liquid-solid (VLS) process then GeSn shells were epitaxially grown around the core NWs (see growth details in Methods). The Sn composition in the shell is controlled by adjusting the growth temperature. This study focuses on six sets of Ge/Ge$_{1-x}$Sn$_{x}$ core/shell NW samples grown at temperatures in the range of 260-310 \degree C. Their composition, structural, electrical, and optical characteristics are described and discussed below.  

The scanning electron microscopy (SEM) image in Fig. 1(a) shows one of the as-grown Ge/Ge$_{1-x}$Sn$_{x}$ core/shell NW samples, where the NW diameter and length are in the range of 100-110 nm and 5-8 $\mu$m, respectively. The inset in Fig. 1(a) displays a representative enlarged SEM image of a single NW. Note that no observable change in the NW morphology is found when the shell growth temperature is varied in the 260-310 \degree C range. The three-dimensional (3-D) atom-by-atom compositional maps of three of these NWs were obtained using atom probe tomography (APT). These analyses yield a Sn content in the inner region of the shells of 8.00$\pm$0.35 at.\%, 14.00$\pm$0.50 at.\%, and 18.00$\pm$0.75 at.\% at a growth temperature of 300 \degree C, 280 \degree C, and 260 \degree C respectively (Fig. 1(b)),  The Sn content of NWs grown at 310 \degree C, 290 \degree C, and 270 \degree C estimated at ~5.6 at.\%, ~10.6 at.\%, and ~15.8 at.\%, respectively. Interestingly, the shell Sn content is found to vary practically linearly with the growth temperature at a rate of 2.5 at.\% per 10 \degree C decrease in temperature. This Sn incorporation rate is higher than that recorded for pseudomorphic (1.3 at.\% /-10\degree C) and relaxed (2.1 at.\% /-10\degree C) thin film growth, as shown in Fig. 1(b).\cite{bouthillier2020decoupling} This behavior confirms the improved strain relaxation and the associated ease of Sn incorporation in the lattice of core/shell NW structures. The high-resolution TEM image of the region in the shell, indicated by the yellow square in the inset of Fig. 1(a), and the corresponding fast Fourier transform (FFT) pattern is displayed in Fig. 1(c). The measured lattice spacing along the [111] direction of 0.328 nm and the sharp diffraction spots in the FFT image are indicative of the good crystalline quality of the shell.

\begin{figure*}[htb]
    \centering
    \includegraphics[width=.8\textwidth]{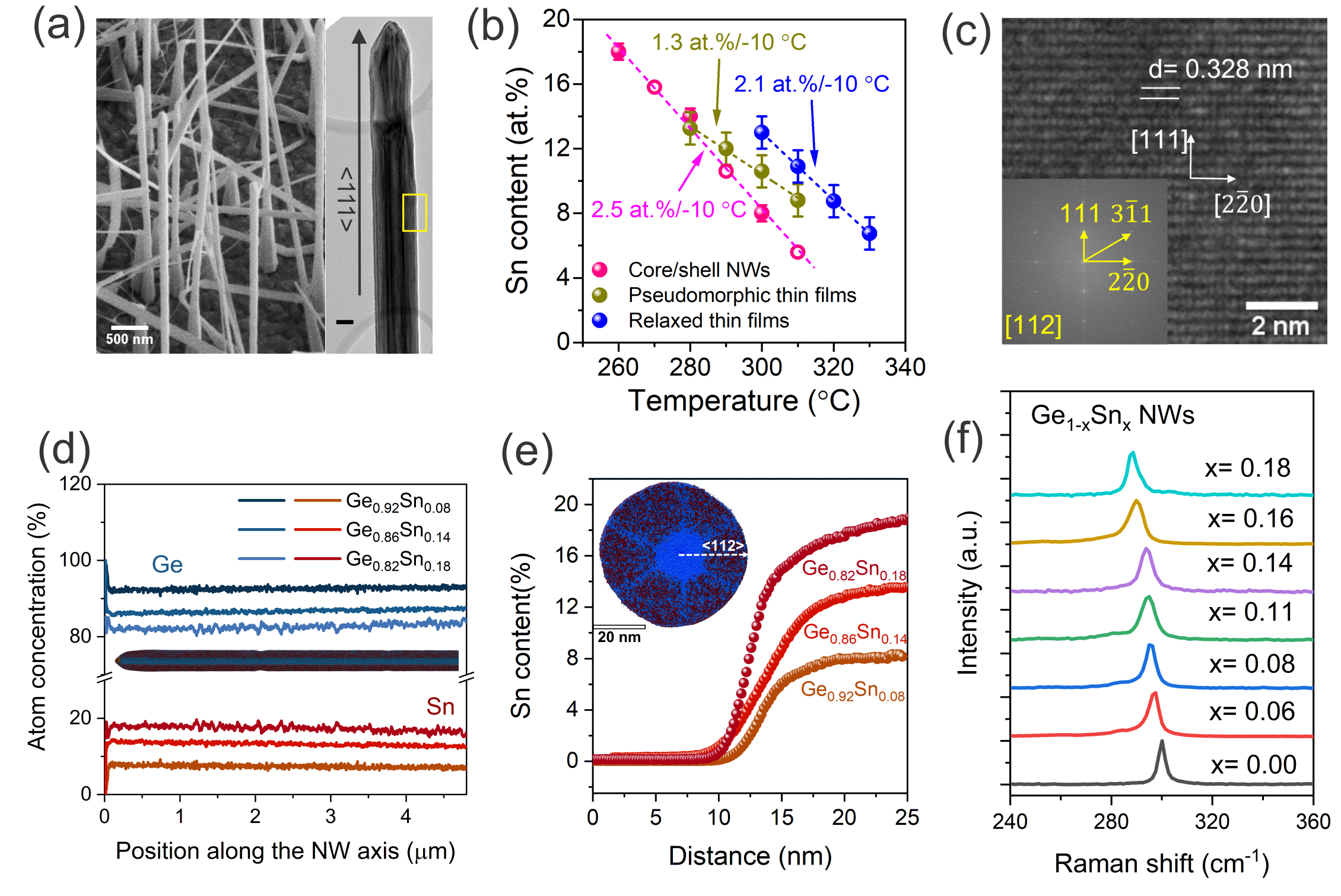}
        \caption{(a) SEM image of the Ge/Ge$_{1-x}$Sn$_{x}$ core/shell NW as-grown sample. The inset shows the zoomed-in SEM image of a single NW, and the scale bar is 50 nm. (b) Sn compositions in core/shell GeSn NWs, pseudomorphic GeSn thin films, and relaxed GeSn thin films as a function of the growth temperature. The pink solid dots represent the data measured by APT, while the pink hollow dots are interpolated values from the growth temperature. Note that the difference in the error bars between NW and thin film content is due to the difference in the characterization techniques; the NW composition was measured by APT, whereas that of thin films by X-ray diffraction (XRD). (c) HR-TEM image of Ge/Ge$_{0.86}$Sn$_{0.14}$ core/shell NW showing the plane direction and the distance of lattice planes. The inset shows the FFT pattern of GeSn shell viewing along the [112] axis. (d) Ge and Sn composition along the growth direction in the shell of the three NWs. (e) Sn composition along the radial direction (from core to shell) in the three NWs. (f) Raman shift of the six Ge/Ge$_{1-x}$Sn$_{x}$ core/shell NWs as well as that of a reference Ge NW.}
    \label{fig.1}
\end{figure*}

Fig. 1(d) and 1(e) show the composition profiles measured by APT for three core/shell NWs grown at 300 \degree C, 280 \degree C, and 260 \degree C. Fig. 1(d) exhibits the composition profile of Ge and Sn elements in the shell extracted along the growth direction for the whole length of the NWs. A marginal change in the Sn content along the NW length is observed between the NW top and bottom regions. Indeed, the difference between the two extremities is around 1.5, 1.0, and 0.7 at.\% for the Ge$_{1-x}$Sn$_{x}$ shells grown at 260 \degree C, 280 \degree C, and 300 \degree C, respectively. These values correspond to a very negligible gradient considering that the NWs are a few micrometers long, thus indicating a high uniformity of the Sn content with an average content of 18.00$\pm$0.75, 14.00$\pm$0.50, and 8.00$\pm$0035 at.\% . The radial Sn composition of three NWs along the $<112>$ direction is displayed in Fig. 1(e), where the inset shows a representative radial atom-by-atom map. These observations show that the use of an ultra-thin Ge core as a growth substrate yields Ge$_{1-x}$Sn$_x$ shells with a homogeneous Sn content as compared to those grown using a thicker Ge core. A comparison of the composition between the core/shell NWs grown with a thin and a thick Ge core can be found in SI.1.

Raman scattering studies were carried out to evaluate the lattice strain and composition in as-grown NWs.\cite{bouthillier2020decoupling} These measurements were performed on individual NWs at room temperature using a 633 nm excitation laser. Fig. 1(f) displays a set of the recorded Raman spectra of a series of six single Ge/Ge$_{1-x}$Sn$_{x}$ core/shell NWs at different contents in addition to a reference spectrum corresponding to that of a single Ge NW. It is noticeable that the Ge-Ge LO mode redshifts from 300.3 cm$^{-1}$ to 296.3 cm$^{-1}$, 294.6 cm$^{-1}$, 293.6 cm$^{-1}$ 292.4 cm$^{-1}$, 290.0 cm$^{-1}$, and 286.7 cm$^{-1}$ as the Sn composition x takes the values of 0.00, 0.06, 0.08, 0.11, 0.14, 0.16, and 0.18, respectively. In addition to the Ge-Ge mode, the Ge-Sn mode around 250-260 cm$^{-1}$ and the disorder-activated (DA)\cite{nelin1972phonon,giannozzi1991ab,bouthillier2020decoupling} mode in 280-290 cm$^{-1}$ range are also detected. A detailed Raman analysis is presented in SI.2. The incorporation of Sn in the Ge lattice causes the Ge-Ge and Ge-Sn modes to downshift because of the relatively larger atomic mass of Sn, while the lattice mismatch-induced compressive strain causes these modes to blueshift. These analyses suggest that the elastic strain in the shell is largely relaxed at the free sidewall facets. Note that the residual strain in the NWs can be estimated by solving the mechanical equilibrium equations with finite element method (FEM) simulation.\cite{albani2018critical,assali2019strain} According to simulations,\cite{albani2018critical,assali2019strain}, the residual strain in the shell decreases with the shell thickness but increases with Sn composition and Ge core diameter. For example, for a Ge core with a diameter of 20 nm, a GeSn shell thickness of 45 nm and a Sn composition of 0.18, the residual compressive strain in the outer shell is below -0.2\%.\cite{albani2018critical} This value, an order of magnitude smaller than that measured in the pseudomorphic thin films, has no observable impact on the measured Raman shift values.  

\textbf{Tunable MWIR photodetection}. Multiple Ge/Ge$_{1-x}$Sn$_{x}$ core/shell NW photoconductive devices were fabricated after transfer on silicon wafer and their cutoff wavelengths were evaluated as a function of Sn content. The fabrication steps are described in the Methods section. A representative SEM image of a multiple NW device is presented in the inset of Fig.2(a). This figure also shows the dark current measured for these devices, where the dotted line is the dark current of a reference device consisting of a Ge$_{0.83}$Sn$_{0.17}$ thin film transferred onto an insulator.\cite{atalla2021all} Compared to the reference device and its Schottky contact, the NW devices display nearly an Ohmic behavior, which is indicative of lower contact resistance in these devices. Note that the dark current of the multiple NW devices is not only dependent on the Sn composition but also on the number of NWs as well as the contact resistance. The number of NWs processed can fluctuate, whereas the contact resistance is related to the density of surface states. Thus, a statistical variation in the dark current was observed among different NW devices at a fixed Sn content, and the data presented in Fig. 2(a) correspond to the average values. 

\begin{figure*}[th]
    \centering
    \includegraphics[width=16cm]{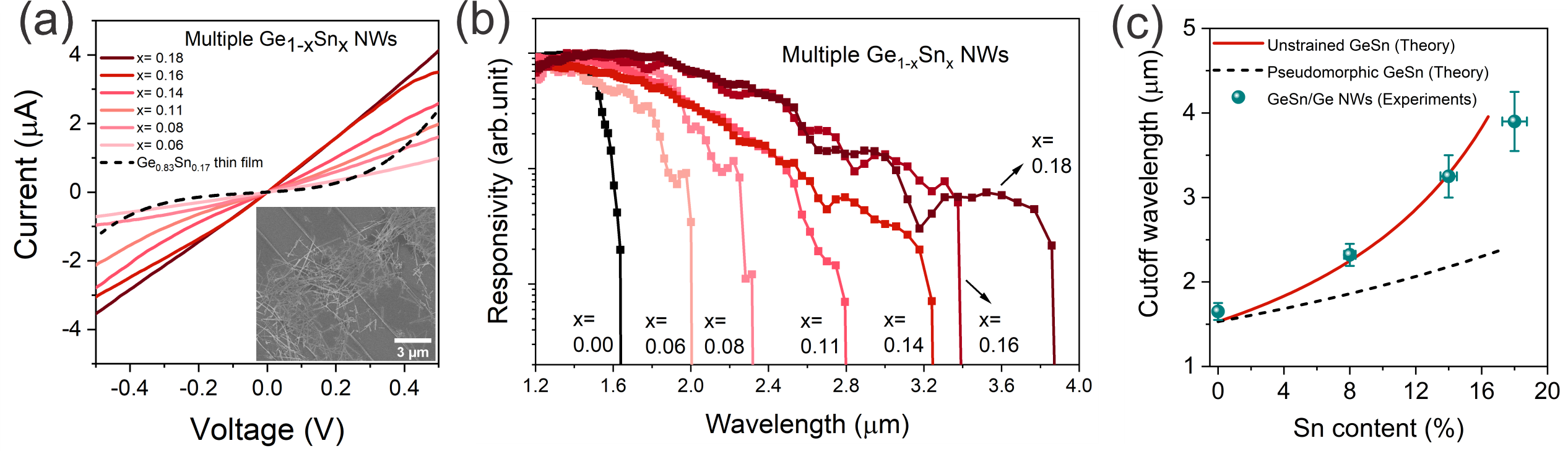}
    \caption{Tunable photodetection from SWIR to MWIR. (a) Dark current of the multiple Ge/Ge$_{1-x}$Sn$_{x}$ core/shell NW devices as well as a transferred Ge$_{0.83}$Sn$_{0.17}$ thin film device. The inset shows the SEM image of a representative multiple NW device. (b) Relative responsivity spectrum of the multiple Ge/Ge$_{1-x}$Sn$_{x}$ core/shell NWs. (c) The calculated and the measured cutoff wavelength as a function of Sn content, where the red solid line and the black dashed line represent the eight-band \kp{} calculation for unstrained and pseudomorphic GeSn alloys, respectively.}
    \label{fig.2}
\end{figure*}

To determine the cutoff wavelength of Ge/Ge$_{1-x}$Sn$_{x}$ core/shell NW devices, room-temperature photocurrent spectrum measurements were conducted. The photocurrent spectra were acquired using a Fourier-transform infrared spectroscopy (FTIR) spectrometer and a lock-in preamplifier (see Methods for details). Fig. 2(b) shows the normalized responsivity as a function of wavelength for six sets of Ge/Ge$_{1-x}$Sn$_{x}$ NW devices as well as that of Ge NW as a reference device. The measured cutoff wavelengths are 1.65 $\mu$m, 2.00 $\mu$m, 2.33 $\mu$m, 2.80 $\mu$m, 3.25 $\mu$m, 3.40 $\mu$m, and 3.85 $\mu$m at a Sn composition of 0.00, 0.06, 0.08, 0.11, 0.14, 0.16 and 0.18, respectively. This result indicates that Ge/Ge$_{1-x}$Sn$_{x}$ core/shell NW photoconductors can cover a large portion of the MWIR by simply tuning the Sn composition. The eight-band \kp{} theory was used to calculate the band gap of the unstrained (red solid line in Fig. 2c) and pseudomorphic (black dashed line in Fig. 2c) Ge$_{1-x}$Sn$_{x}$ alloys. It is evident that the pseudomorphic GeSn exhibits a shorter cutoff wavelength compared to unstrained GeSn and the difference between the two values increases with Sn content. This means that compressive strain in GeSn alloys increases the band gap besides limiting its directness. However, as displayed in Fig. 2(c), the measured cutoff wavelengths of Ge NW and Ge/Ge$_{1-x}$Sn$_{x}$ core/shell NWs at x = 0.08, 0.14, and 0.18 show good agreement with the calculation of unstrained GeSn alloys, which provides an additional confirmation of the strain relaxation in the NWs. Note that the eight-band \kp{} framework is based on the parameterization reported in the literature\cite{Changkp2010,LuLowkp2012,Polak_2017}. To account for the inaccuracy of Vegard's law to estimate the bandgap of \GeSn{1-x}{x} alloys, bandgap bowing parameters are introduced for L and $\Gamma$ high-symmetry points. In these calculations, only \GeSn{1-x}{x} alloys with Sn composition of 17\% and lower are considered in the bandgap calculation because of the lack of reliable material parameters for alloys above this content.\cite{rainko2018investigation,moutanabbir2021monolithic} Additionally, the impact of the substrate crystallographic orientation is taken into account by introducing different rotation matrices in the calculation of the Hamiltonian.\cite{Eisfeller2012} The fact that the measured values are consistent with the calculated bandgap energies for fully relaxed alloys is indicative of the enhanced strain relaxation in the Ge$_{1-x}$Sn$_{x}$ shells grown on thin Ge NWs. 

\textbf{Photocurrent at 1550 nm and 2330 nm excitations}. The photoresponse of Ge/Ge$_{0.86}$Sn$_{0.14}$ core/shell NW devices was investigated under a 1550 nm laser and a 2330 nm laser without a lock-in preamplifier (Fig. 2).
\begin{figure*}[th]
    \centering
    \includegraphics[width=16cm]{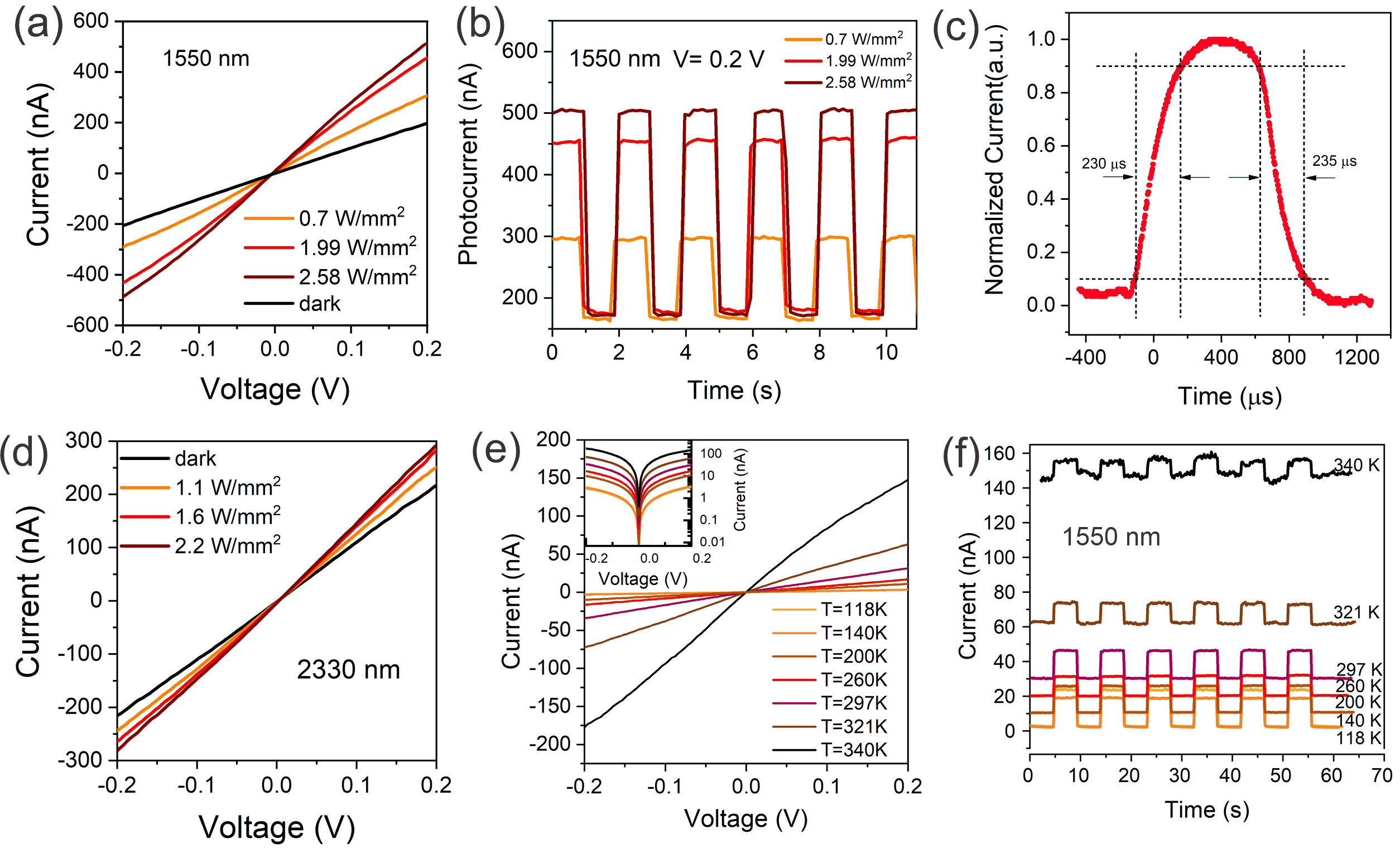}
    \caption{Photocurrent of a multiple Ge/Ge$_{0.86}$Sn$_{0.14}$ core/shell NW device. (a) Dark current and photocurrent with different power densities under a 1550 nm laser and (d) a 2330 nm laser. (b) The transient photocurrent of the multiple NW device under the 1550 nm laser with different power densities. (c) High-resolution transient photoresponse of the multiple NW device to illustrate the rise and fall time under the 1550 nm laser. (e) The temperature-dependent dark current of the multiple Ge/Ge$_{0.86}$Sn$_{0.14}$ core/shell NW device. (f) The temperature-dependent photocurrent of the multiple Ge/Ge$_{0.86}$Sn$_{0.14}$ core/shell NW device at 1550 nm.}
    \label{fig.3}
\end{figure*}
Fig. 3(a) and (d) show the dark current and the photocurrent under two excitation wavelengths at variable power densities. The transient photocurrent under 1550 nm illumination is shown in Fig. 3(b), and the transient photocurrent under 2330 nm illumination is provided in Fig. S3(a). The incident light is switched with an on/off interval of 1 s. Fig. 3(b) presents the transient photocurrent during six repeated cycles, showing a consistent and repeatable photocurrent during the measurement. The rise time, which is the time to increase from 10\% to 90\% of the maximum photocurrent, and the fall time, which is the time to recover from 90\% to 10\% of the maximum photocurrent, are 230 $\mu$s and 235 $\mu$s at 1550 nm, respectively (Fig. 3(c)). The measured rise and fall times under the 2330 nm illumination are 265 $\mu$s and 280 $\mu$s, respectively  (Fig. S3). The photocurrent obtained under 1550 nm illumination and a bias of 0.2 V with a power density of 1.99 W/mm$^2$ is 275 nA, which is fourfold higher than that measured under the 2330 nm illumination. This observation aligns well with the responsivity spectrum displayed in Fig. 2(b). 

Since the dark current in a photoconductive device is sensitive to temperature, it is necessary to also investigate the device performance under variable temperatures. Fig. 3(e) shows the dark current of a Ge/Ge$_{0.86}$Sn$_{0.14}$ multiple NW device at different temperatures ranging from 118 K to 340 K (see Methods for details). The dark current is remarkably reduced as temperature decreases most likely because the intrinsic excitation is suppressed in low temperatures. For instance, the dark current at 340 K is $\sim$ 75 X higher than that at 118 K. The inset shows the temperature-dependent dark current with a semi-logarithmic scale. The temperature-dependent photocurrent measurement under 1550 nm illumination is displayed in Fig. 3(f), and it is evident that the photocurrent slightly increases from 9 nA to 22 nA at 0.2 V bias as the temperature decreases from 340 K to 118 K.  The weak enhanced light response is likely
due to the suppressed noise current and enhanced absorption. 

\textbf{NW single-pixel infrared imaging}. The obtained NW devices were subsequently integrated into infrared imaging systems. As illustrated in Fig. S5, the major components of the imaging setup include a light source, a focus lens, two reflective mirrors, and a motorized XY stage. A broadband near-infrared lamp (0.3-2.6 $\mu$m), and two single-wavelength lasers (1.55 $\mu$m and 2.33 $\mu$m) were used as illumination sources. A 3D-printed maple leaf pattern with an area of 1.2 cm$\times$1.5 cm and a lily flower pattern with an area of 1.0 cm$\times$1.8 cm were installed on the motorized XY stage. The first two rows (Fig. 4(a) and (b)) are the maple leaf and lily flower images acquired under the broadband near-infrared light source with and without an 800 nm long-pass filter. Fig. 4(c) and (d) exhibit the two images recorded under 1.55 $\mu$m  and 2.33 $\mu$m illuminations, respectively. 
\begin{figure*}[th]
    \centering
    \includegraphics[width=16cm]{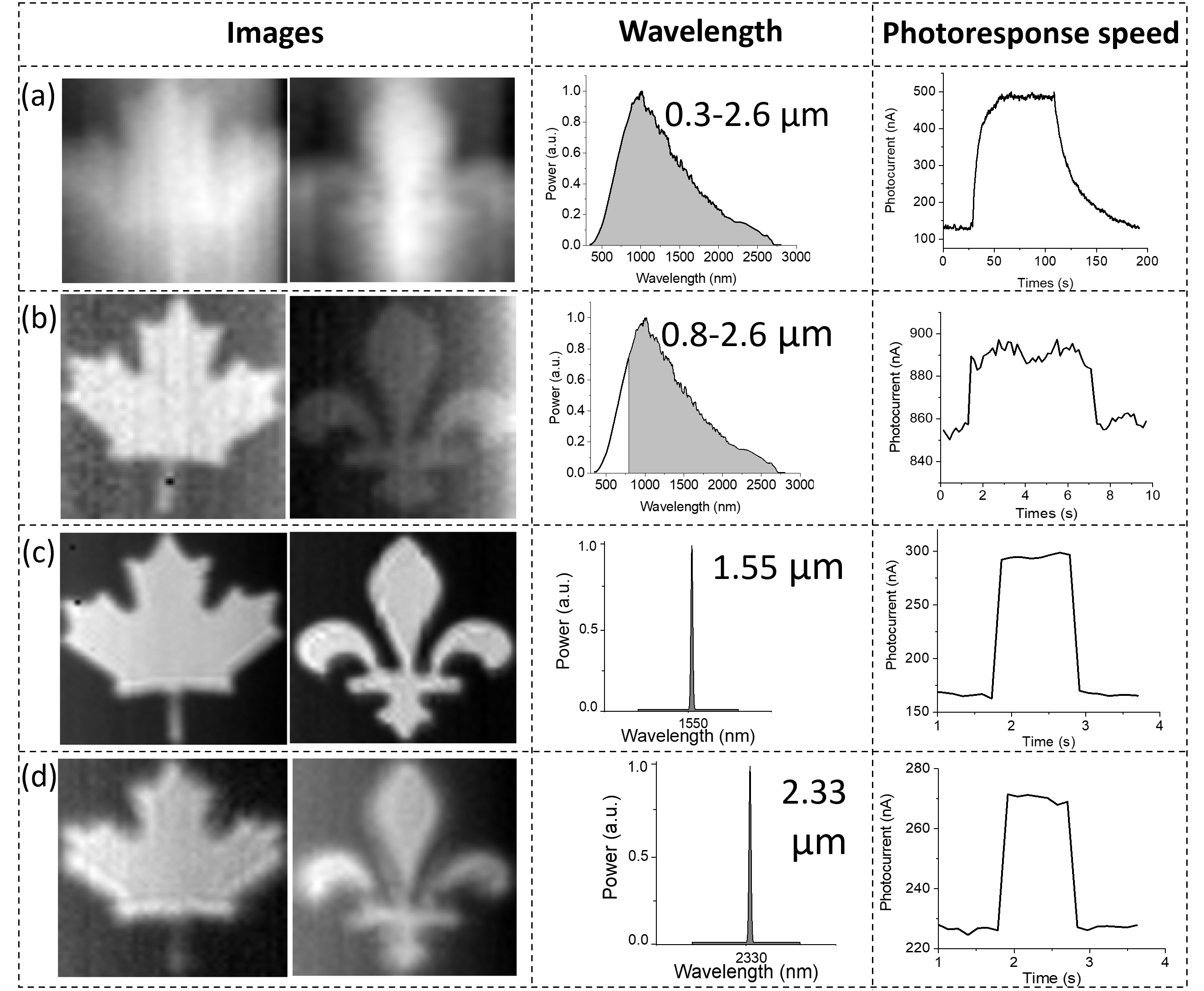}
    \caption{NW single-pixel imaging measurement. (a) Measured maple leaf and lily flower images under illumination in the range of 0.3-2.6 $\mu$m and the corresponding photoresponse speed. (b) Measured maple leaf and lily flower images under illumination in the range of 0.8-2.6 $\mu$m and the corresponding photoresponse speed. (c) and (d) Measured maple leaf and lily flower images and the corresponding photoresponse speed at 1.55 $\mu$m and at 2.33 $\mu$m.}
    \label{fig.4}
\end{figure*}
The maple leaf image is composed of 60x60 pixels while the lily flower image contains 50x65 pixels. It is noticeable that the images in Fig. 4(a) obtained under the broadband illumination are blurry, but their quality is improved when the visible light is suppressed by adding an 800 nm long-pass filter, as shown in Fig. 4(b). In fact, the image quality is mainly determined by the photoresponse speed of the NW photodetectors under each illumination. For example, the images in Fig. 4(c) and 4(d) are clear because the response speeds of the NW device under the two wavelengths are faster than the movement speed of the XY motorized stage. However, the rise and fall times of the NW photodetector under illumination in the range of 0.3-2.6 $\mu$m is very long (> 50 s) compared to the step delay of the motorized XY stage (1.5 s). That means the current change in the NW detector is unable to keep pace with the stage movement, therefore inducing blurred images.

The surface states in NWs can act as trap centers for the photogenerated carriers, therefore prolonging the carrier lifetime,\cite{yamamoto2018pronounced} which is broadly known as the photogating effect.\cite{fang2017photogating} The photogating effect is a common phenomenon in low-dimensional photoconductive detectors when they are exposed to visible light. However, this effect is largely reduced when the visible light is cut off, thus leading to an enhanced image quality in Fig. 4(b). Note that the dark current of Ge/Ge$_{1-x}$Sn$_{x}$ core/shell NW devices is sensitive to ambient thermal radiation as well as to illumination-induced NW heating. The latter becomes prevalent when the measurement is extended over a long period reaching hours, especially under the 2330 nm laser. Therefore, the dark current can slowly increase during the long imaging process, leading to a slightly brighter background on one side of the images. Moreover, the image quality of the lily flower in Fig. 4(d) is comparatively lower due to the relatively unstable contact between the probe and the device.

\section{Discussion}
The tunable cutoff wavelength, the fast and stable photoresponse, and the good image quality recorded without cooling indicate that Ge/Ge$_{1-x}$Sn$_{x}$ core/shell NWs could be an effective platform for silicon-compatible, monolithic MWIR devices. It is indeed rather straightforward to narrow the bandgap by just controlling the Sn composition in Ge$_{1-x}$Sn$_x$ alloys exploiting the compliant nature of sub-20 nm Ge NWs. Note that the demonstrated room-temperature imaging was achieved by the NW photoconductor devices without using a lock-in preamplifier, which is an advantage \textit{per se} rooted in the relatively low dark current and high responsivity of these NW devices. Indeed, both cooling and lock-in techniques are required for imaging using GeSn thin film photoconductors.\cite{tran2019si} Recent studies showed that GeSn thin layer PIN photodiodes show clear improvement,\cite{talamas2021cmos} thus it is reasonable to envision an enhanced imaging performance of NW-based PIN detectors. The measured cutoff wavelength generally showed good agreement with the calculated values from the eight-band \kp{} theory when the Sn content is below 17 at.\%, however, the impact of atomistic effects on the electronic structure evolution in GeSn alloys should be considered in order to increase the accuracy of these calculations.\cite{moutanabbir2021monolithic} In this work, the multiple Ge/GeSn core/shell NW devices allow for a more straightforward probe of the bandgap energy using photocurrent spectral measurements as compared to single NW devices. Both the small volume of the material in the single NW device and the limited power of the light source, especially in the MWIR range, can limit the generated photocurrent thereby underestimating the cutoff wavelength. The transient photoresponse of the NW devices under illumination using two different lasers indicates a reproducible and stable photodetection performance. The 3-dB bandwidth $f_{3dB}$ of the multiple NW photodetector, obtained using:\cite{jie2006photoconductive} $f_{3dB} = 0.35/\tau_{rise}$, is 1.52 kHz at 1550 nm and 1.32 kHz at 2330 nm. The response speed of a photoconductor device is limited by the long carrier lifetime, which can be affected by the high density of carrier traps in NWs due to the large surface-to-volume ratio.\cite{zheng2018high,wu2011fast} In fact, it is typical for a low-dimensional photoconductive detector with an obvious photogating phenomenon to have a bandwidth that is lower than 1 MHz.\cite{fang2017photogating} Ideally, in the single-pixel imaging measurement, when the step delay is set to 1.5 s, it would take about 1.5 hours to record an image of 1.2 cm$\times$1.5 cm in size. However, the actual measurement time is approximately 4 hours. The 2.5-hour delay is attributed to the communication lag between the Labview software and the external hardware. Therefore, if there is no software and hardware communication delay, and the step delay is set as the response time of the NW device, the imaging time of an identical pattern will be reduced to a few minutes. 

\section{CONCLUSION}\label{sec:concl}
In summary, sub-20 nm Ge NWs have proven to be effective compliant nanoscale substrates to grow Ge$_{1-x}$Sn$_{x}$ alloys with remarkable Sn content uniformity over micrometer scales without a significant compressive strain build-up in their lattice for ${x}$ between 0.06 and 0.18. Multiple Ge/Ge$_{1-x}$Sn$_{x}$ core/shell NW photoconductive devices were also fabricated exhibiting relatively high and fast photoresponse at excitation wavelengths of 1550 nm and 2330 nm at room temperature. These detectors show a tunable cutoff wavelength covering the 2.0 $\mu$m to 3.9 $\mu$m range. A slight enhancement in the photoresponse performance was observed when operating at low temperatures. Moreover, the obtained photoconductors were integrated into uncooled imagers enabling the acquisition of high-quality images under a broadband lamp and 1550 nm and 2330 nm lasers without the lock-in technique. These results highlight the potential of Ge/Ge$_{1-x}$Sn$_{x}$ core/shell NWs as a promising material for scalable, compact, and cost-effective MWIR photonic and optoelectronic devices.\\

\section{METHODS}\label{sec:method}
\textbf{Growth of Ge/Ge$_{1-x}$Sn$_{x}$ core/shell NWs} Ge/Ge$_{1-x}$Sn$_{x}$ core/shell NWs were grown on Ge (111) substrates in a CVD reactor following a two-step growth protocol. First, Ge NWs were synthesized at 340 \degree C by the VLS process, where monogermane (GeH$_4$) and 20 nm gold droplets served as the growth precursor and catalyst, respectively. Second, Ge$_{1-x}$Sn$_{x}$ shell layers of 40-50 nm in thickness were epitaxially grown around the Ge core by supplying GeH$_4$ and tetrachloride tin (SnCl$_4$) precursors. The growth temperature in the second step was adjusted to control the Sn content. Six Ge/Ge$_{1-x}$Sn$_{x}$ core/shell sets of NWs, grown at 310 \degree C, 300 \degree C, 290 \degree C, 280 \degree C, 270 \degree C, and 260 \degree C, were investigated in this study.\\
\textbf{Materials characterization}
TEM and APT specimens were prepared in a Thermo Fisher Helios Nanolab 660 dual-beam scanning electron microscope using a gallium-focused ion beam (FIB) at 30, 16, and 5~kV. For the TEM analyses, electron beam-induced carbon and platinum were locally deposited on the sample as a contrast layer in addition to a layer of ion beam-deposited platinum to protect the imaged region from any damage by the ion-beam milling during the thinning of the TEM lamella. TEM analyses were carried out on a Thermo Scientific Talos F200X~S/TEM system with an acceleration voltage of 200~kV. For APT analyses, single nanowires were transferred onto silicon pedestals using a method described in Ref.\cite{koelling2017atom} Laser-assisted atom probe measurements were performed in a LEAP~5000~XS utilizing a picosecond laser to generate pulses at a wavelength of 355~nm. For the analysis, all samples were cooled to a temperature of 28~K. The experimental data were collected at laser pulse energies of 3-7.5~pJ.\\
\textbf{Device fabrication and photocurrent analysis} Interdigitated contacts were fabricated by a mask aligner after transferring the multiple NWs on a SiO$_2$/Si substrate. Subsequently, a standard lift-off process was performed following the E-beam evaporation of Cr/Au (5/150 nm). The photocurrent spectrum was measured with a Bruker Vertex 80 Fourier-transform infrared spectroscopy (FTIR) spectrometer connected with a Zurich Instruments lock-in amplifier which is triggered by an optical chopper. A broadband light source chopped by the optical chopper was focused by a 40× objective lens before being sent to the NW devices. The electrical signal was sent to the lock-in amplifier before being fed into FTIR to get the photocurrent spectrum. The power of the broadband light source was calibrated by a commercial InGaAs photodiode.\\
\textbf{Temperature-dependent dark current and photocurrent analyses} A cryogenic probe station from Janis Research Company was used to perform the temperature-dependent measurement and the light from a 1550 nm butterfly laser diode was sent to the device through an optical fiber, which created a 50 $\mu$m light spot on the samples. The chamber was evacuated by a mechanical pump, and the liquid Ni was sent to the chamber after the pressure reached 10$^{-8}$ Psi. The temperature in the chamber was controlled by a PID controller. 

\medskip
\noindent {\textbf{ACKNOWLEDGEMENTS}}.
The authors thank J. Bouchard for the technical support with the CVD system, Nathalie Rojas-Lobo for providing the 3D-printed patterns, and Yves-Alain Peter and C\'edric Lemieux-Leduc for providing the oscilloscope.  O.M. acknowledges support from NSERC Canada (Discovery, SPG, and CRD Grants), Canada Research Chairs, Canada Foundation for Innovation, Mitacs, PRIMA Québec, and Defence Canada (Innovation for Defence Excellence and Security, IDEaS). L.L acknowledges support from China Scholarship Council (CSC). \\

\medskip
\noindent {\bf COMPETING INTERESTS}.
The authors declare no competing financial interests.

\medskip
\noindent {\bf DATA AVAILABILITY}.
The data that support the findings of this study are available from the corresponding authors upon reasonable request.
\bigskip
\bibliography{NW_main.bib} 

\begin{thebibliography}{54}%
\makeatletter
\providecommand \@ifxundefined [1]{%
 \@ifx{#1\undefined}
}%
\providecommand \@ifnum [1]{%
 \ifnum #1\expandafter \@firstoftwo
 \else \expandafter \@secondoftwo
 \fi
}%
\providecommand \@ifx [1]{%
 \ifx #1\expandafter \@firstoftwo
 \else \expandafter \@secondoftwo
 \fi
}%
\providecommand \natexlab [1]{#1}%
\providecommand \enquote  [1]{``#1''}%
\providecommand \bibnamefont  [1]{#1}%
\providecommand \bibfnamefont [1]{#1}%
\providecommand \citenamefont [1]{#1}%
\providecommand \href@noop [0]{\@secondoftwo}%
\providecommand \href [0]{\begingroup \@sanitize@url \@href}%
\providecommand \@href[1]{\@@startlink{#1}\@@href}%
\providecommand \@@href[1]{\endgroup#1\@@endlink}%
\providecommand \@sanitize@url [0]{\catcode `\\12\catcode `\$12\catcode `\&12\catcode `\#12\catcode `\^12\catcode `\_12\catcode `\%12\relax}%
\providecommand \@@startlink[1]{}%
\providecommand \@@endlink[0]{}%
\providecommand \url  [0]{\begingroup\@sanitize@url \@url }%
\providecommand \@url [1]{\endgroup\@href {#1}{\urlprefix }}%
\providecommand \urlprefix  [0]{URL }%
\providecommand \Eprint [0]{\href }%
\providecommand \doibase [0]{https://doi.org/}%
\providecommand \selectlanguage [0]{\@gobble}%
\providecommand \bibinfo  [0]{\@secondoftwo}%
\providecommand \bibfield  [0]{\@secondoftwo}%
\providecommand \translation [1]{[#1]}%
\providecommand \BibitemOpen [0]{}%
\providecommand \bibitemStop [0]{}%
\providecommand \bibitemNoStop [0]{.\EOS\space}%
\providecommand \EOS [0]{\spacefactor3000\relax}%
\providecommand \BibitemShut  [1]{\csname bibitem#1\endcsname}%
\let\auto@bib@innerbib\@empty
\bibitem [{\citenamefont {Moutanabbir}\ \emph {et~al.}(2021)\citenamefont {Moutanabbir}, \citenamefont {Assali}, \citenamefont {Gong}, \citenamefont {O'Reilly}, \citenamefont {Broderick}, \citenamefont {Marzban}, \citenamefont {Witzens}, \citenamefont {Du}, \citenamefont {Yu}, \citenamefont {Chelnokov} \emph {et~al.}}]{moutanabbir2021monolithic}%
  \BibitemOpen
  \bibfield  {author} {\bibinfo {author} {\bibfnamefont {O.}~\bibnamefont {Moutanabbir}}, \bibinfo {author} {\bibfnamefont {S.}~\bibnamefont {Assali}}, \bibinfo {author} {\bibfnamefont {X.}~\bibnamefont {Gong}}, \bibinfo {author} {\bibfnamefont {E.}~\bibnamefont {O'Reilly}}, \bibinfo {author} {\bibfnamefont {C.}~\bibnamefont {Broderick}}, \bibinfo {author} {\bibfnamefont {B.}~\bibnamefont {Marzban}}, \bibinfo {author} {\bibfnamefont {J.}~\bibnamefont {Witzens}}, \bibinfo {author} {\bibfnamefont {W.}~\bibnamefont {Du}}, \bibinfo {author} {\bibfnamefont {S.-Q.}\ \bibnamefont {Yu}}, \bibinfo {author} {\bibfnamefont {A.}~\bibnamefont {Chelnokov}}, \emph {et~al.},\ }\href@noop {} {\bibfield  {journal} {\bibinfo  {journal} {Applied Physics Letters}\ }\textbf {\bibinfo {volume} {118}},\ \bibinfo {pages} {110502} (\bibinfo {year} {2021})}\BibitemShut {NoStop}%
\bibitem [{\citenamefont {Buca}\ \emph {et~al.}(2022)\citenamefont {Buca}, \citenamefont {Bjelajac}, \citenamefont {Spirito}, \citenamefont {Concepci{\'o}n}, \citenamefont {Gromovyi}, \citenamefont {Sakat}, \citenamefont {Lafosse}, \citenamefont {Ferlazzo}, \citenamefont {von~den Driesch}, \citenamefont {Ikonic} \emph {et~al.}}]{buca2022room}%
  \BibitemOpen
  \bibfield  {author} {\bibinfo {author} {\bibfnamefont {D.}~\bibnamefont {Buca}}, \bibinfo {author} {\bibfnamefont {A.}~\bibnamefont {Bjelajac}}, \bibinfo {author} {\bibfnamefont {D.}~\bibnamefont {Spirito}}, \bibinfo {author} {\bibfnamefont {O.}~\bibnamefont {Concepci{\'o}n}}, \bibinfo {author} {\bibfnamefont {M.}~\bibnamefont {Gromovyi}}, \bibinfo {author} {\bibfnamefont {E.}~\bibnamefont {Sakat}}, \bibinfo {author} {\bibfnamefont {X.}~\bibnamefont {Lafosse}}, \bibinfo {author} {\bibfnamefont {L.}~\bibnamefont {Ferlazzo}}, \bibinfo {author} {\bibfnamefont {N.}~\bibnamefont {von~den Driesch}}, \bibinfo {author} {\bibfnamefont {Z.}~\bibnamefont {Ikonic}}, \emph {et~al.},\ }\href@noop {} {\bibfield  {journal} {\bibinfo  {journal} {Advanced Optical Materials}\ }\textbf {\bibinfo {volume} {10}},\ \bibinfo {pages} {2201024} (\bibinfo {year} {2022})}\BibitemShut {NoStop}%
\bibitem [{\citenamefont {Atalla}\ \emph {et~al.}(2023{\natexlab{a}})\citenamefont {Atalla}, \citenamefont {Kim}, \citenamefont {Assali}, \citenamefont {Burt}, \citenamefont {Nam},\ and\ \citenamefont {Moutanabbir}}]{atalla2023extended}%
  \BibitemOpen
  \bibfield  {author} {\bibinfo {author} {\bibfnamefont {M.~R.}\ \bibnamefont {Atalla}}, \bibinfo {author} {\bibfnamefont {Y.}~\bibnamefont {Kim}}, \bibinfo {author} {\bibfnamefont {S.}~\bibnamefont {Assali}}, \bibinfo {author} {\bibfnamefont {D.}~\bibnamefont {Burt}}, \bibinfo {author} {\bibfnamefont {D.}~\bibnamefont {Nam}},\ and\ \bibinfo {author} {\bibfnamefont {O.}~\bibnamefont {Moutanabbir}},\ }\href@noop {} {\bibfield  {journal} {\bibinfo  {journal} {ACS Photonics}\ }\textbf {\bibinfo {volume} {10}},\ \bibinfo {pages} {1649} (\bibinfo {year} {2023}{\natexlab{a}})}\BibitemShut {NoStop}%
\bibitem [{\citenamefont {Chang}\ \emph {et~al.}(2022)\citenamefont {Chang}, \citenamefont {Yeh}, \citenamefont {Jheng}, \citenamefont {Hsu}, \citenamefont {Lee}, \citenamefont {Li}, \citenamefont {Cheng},\ and\ \citenamefont {Chang}}]{chang2022mid}%
  \BibitemOpen
  \bibfield  {author} {\bibinfo {author} {\bibfnamefont {C.-Y.}\ \bibnamefont {Chang}}, \bibinfo {author} {\bibfnamefont {P.-L.}\ \bibnamefont {Yeh}}, \bibinfo {author} {\bibfnamefont {Y.-T.}\ \bibnamefont {Jheng}}, \bibinfo {author} {\bibfnamefont {L.-Y.}\ \bibnamefont {Hsu}}, \bibinfo {author} {\bibfnamefont {K.-C.}\ \bibnamefont {Lee}}, \bibinfo {author} {\bibfnamefont {H.}~\bibnamefont {Li}}, \bibinfo {author} {\bibfnamefont {H.}~\bibnamefont {Cheng}},\ and\ \bibinfo {author} {\bibfnamefont {G.-E.}\ \bibnamefont {Chang}},\ }\href@noop {} {\bibfield  {journal} {\bibinfo  {journal} {Photonics Research}\ }\textbf {\bibinfo {volume} {10}},\ \bibinfo {pages} {2278} (\bibinfo {year} {2022})}\BibitemShut {NoStop}%
\bibitem [{\citenamefont {Chretien}\ \emph {et~al.}(2019)\citenamefont {Chretien}, \citenamefont {Pauc}, \citenamefont {Armand~Pilon}, \citenamefont {Bertrand}, \citenamefont {Thai}, \citenamefont {Casiez}, \citenamefont {Bernier}, \citenamefont {Dansas}, \citenamefont {Gergaud}, \citenamefont {Delamadeleine} \emph {et~al.}}]{chretien2019gesn}%
  \BibitemOpen
  \bibfield  {author} {\bibinfo {author} {\bibfnamefont {J.}~\bibnamefont {Chretien}}, \bibinfo {author} {\bibfnamefont {N.}~\bibnamefont {Pauc}}, \bibinfo {author} {\bibfnamefont {F.}~\bibnamefont {Armand~Pilon}}, \bibinfo {author} {\bibfnamefont {M.}~\bibnamefont {Bertrand}}, \bibinfo {author} {\bibfnamefont {Q.-M.}\ \bibnamefont {Thai}}, \bibinfo {author} {\bibfnamefont {L.}~\bibnamefont {Casiez}}, \bibinfo {author} {\bibfnamefont {N.}~\bibnamefont {Bernier}}, \bibinfo {author} {\bibfnamefont {H.}~\bibnamefont {Dansas}}, \bibinfo {author} {\bibfnamefont {P.}~\bibnamefont {Gergaud}}, \bibinfo {author} {\bibfnamefont {E.}~\bibnamefont {Delamadeleine}}, \emph {et~al.},\ }\href@noop {} {\bibfield  {journal} {\bibinfo  {journal} {ACS Photonics}\ }\textbf {\bibinfo {volume} {6}},\ \bibinfo {pages} {2462} (\bibinfo {year} {2019})}\BibitemShut {NoStop}%
\bibitem [{\citenamefont {Chr{\'e}tien}\ \emph {et~al.}(2022)\citenamefont {Chr{\'e}tien}, \citenamefont {Thai}, \citenamefont {Frauenrath}, \citenamefont {Casiez}, \citenamefont {Chelnokov}, \citenamefont {Reboud}, \citenamefont {Hartmann}, \citenamefont {El~Kurdi}, \citenamefont {Pauc},\ and\ \citenamefont {Calvo}}]{chretien2022room}%
  \BibitemOpen
  \bibfield  {author} {\bibinfo {author} {\bibfnamefont {J.}~\bibnamefont {Chr{\'e}tien}}, \bibinfo {author} {\bibfnamefont {Q.}~\bibnamefont {Thai}}, \bibinfo {author} {\bibfnamefont {M.}~\bibnamefont {Frauenrath}}, \bibinfo {author} {\bibfnamefont {L.}~\bibnamefont {Casiez}}, \bibinfo {author} {\bibfnamefont {A.}~\bibnamefont {Chelnokov}}, \bibinfo {author} {\bibfnamefont {V.}~\bibnamefont {Reboud}}, \bibinfo {author} {\bibfnamefont {J.}~\bibnamefont {Hartmann}}, \bibinfo {author} {\bibfnamefont {M.}~\bibnamefont {El~Kurdi}}, \bibinfo {author} {\bibfnamefont {N.}~\bibnamefont {Pauc}},\ and\ \bibinfo {author} {\bibfnamefont {V.}~\bibnamefont {Calvo}},\ }\href@noop {} {\bibfield  {journal} {\bibinfo  {journal} {Applied Physics Letters}\ }\textbf {\bibinfo {volume} {120}},\ \bibinfo {pages} {051107} (\bibinfo {year} {2022})}\BibitemShut {NoStop}%
\bibitem [{\citenamefont {Atalla}\ \emph {et~al.}(2021)\citenamefont {Atalla}, \citenamefont {Assali}, \citenamefont {Attiaoui}, \citenamefont {Lemieux-Leduc}, \citenamefont {Kumar}, \citenamefont {Abdi},\ and\ \citenamefont {Moutanabbir}}]{atalla2021all}%
  \BibitemOpen
  \bibfield  {author} {\bibinfo {author} {\bibfnamefont {M.~R.}\ \bibnamefont {Atalla}}, \bibinfo {author} {\bibfnamefont {S.}~\bibnamefont {Assali}}, \bibinfo {author} {\bibfnamefont {A.}~\bibnamefont {Attiaoui}}, \bibinfo {author} {\bibfnamefont {C.}~\bibnamefont {Lemieux-Leduc}}, \bibinfo {author} {\bibfnamefont {A.}~\bibnamefont {Kumar}}, \bibinfo {author} {\bibfnamefont {S.}~\bibnamefont {Abdi}},\ and\ \bibinfo {author} {\bibfnamefont {O.}~\bibnamefont {Moutanabbir}},\ }\href@noop {} {\bibfield  {journal} {\bibinfo  {journal} {Advanced Functional Materials}\ }\textbf {\bibinfo {volume} {31}},\ \bibinfo {pages} {2006329} (\bibinfo {year} {2021})}\BibitemShut {NoStop}%
\bibitem [{\citenamefont {Atalla}\ \emph {et~al.}(2022)\citenamefont {Atalla}, \citenamefont {Assali}, \citenamefont {Koelling}, \citenamefont {Attiaoui},\ and\ \citenamefont {Moutanabbir}}]{atalla2022high}%
  \BibitemOpen
  \bibfield  {author} {\bibinfo {author} {\bibfnamefont {M.~R.}\ \bibnamefont {Atalla}}, \bibinfo {author} {\bibfnamefont {S.}~\bibnamefont {Assali}}, \bibinfo {author} {\bibfnamefont {S.}~\bibnamefont {Koelling}}, \bibinfo {author} {\bibfnamefont {A.}~\bibnamefont {Attiaoui}},\ and\ \bibinfo {author} {\bibfnamefont {O.}~\bibnamefont {Moutanabbir}},\ }\href@noop {} {\bibfield  {journal} {\bibinfo  {journal} {ACS Photonics}\ }\textbf {\bibinfo {volume} {9}},\ \bibinfo {pages} {1425} (\bibinfo {year} {2022})}\BibitemShut {NoStop}%
\bibitem [{\citenamefont {Elbaz}\ \emph {et~al.}(2020{\natexlab{a}})\citenamefont {Elbaz}, \citenamefont {Buca}, \citenamefont {von~den Driesch}, \citenamefont {Pantzas}, \citenamefont {Patriarche}, \citenamefont {Zerounian}, \citenamefont {Herth}, \citenamefont {Checoury}, \citenamefont {Sauvage}, \citenamefont {Sagnes} \emph {et~al.}}]{elbaz2020ultra}%
  \BibitemOpen
  \bibfield  {author} {\bibinfo {author} {\bibfnamefont {A.}~\bibnamefont {Elbaz}}, \bibinfo {author} {\bibfnamefont {D.}~\bibnamefont {Buca}}, \bibinfo {author} {\bibfnamefont {N.}~\bibnamefont {von~den Driesch}}, \bibinfo {author} {\bibfnamefont {K.}~\bibnamefont {Pantzas}}, \bibinfo {author} {\bibfnamefont {G.}~\bibnamefont {Patriarche}}, \bibinfo {author} {\bibfnamefont {N.}~\bibnamefont {Zerounian}}, \bibinfo {author} {\bibfnamefont {E.}~\bibnamefont {Herth}}, \bibinfo {author} {\bibfnamefont {X.}~\bibnamefont {Checoury}}, \bibinfo {author} {\bibfnamefont {S.}~\bibnamefont {Sauvage}}, \bibinfo {author} {\bibfnamefont {I.}~\bibnamefont {Sagnes}}, \emph {et~al.},\ }\href@noop {} {\bibfield  {journal} {\bibinfo  {journal} {Nature Photonics}\ }\textbf {\bibinfo {volume} {14}},\ \bibinfo {pages} {375} (\bibinfo {year} {2020}{\natexlab{a}})}\BibitemShut {NoStop}%
\bibitem [{\citenamefont {Joo}\ \emph {et~al.}(2021)\citenamefont {Joo}, \citenamefont {Kim}, \citenamefont {Burt}, \citenamefont {Jung}, \citenamefont {Zhang}, \citenamefont {Chen}, \citenamefont {Parluhutan}, \citenamefont {Kang}, \citenamefont {Lee}, \citenamefont {Assali} \emph {et~al.}}]{joo20211d}%
  \BibitemOpen
  \bibfield  {author} {\bibinfo {author} {\bibfnamefont {H.-J.}\ \bibnamefont {Joo}}, \bibinfo {author} {\bibfnamefont {Y.}~\bibnamefont {Kim}}, \bibinfo {author} {\bibfnamefont {D.}~\bibnamefont {Burt}}, \bibinfo {author} {\bibfnamefont {Y.}~\bibnamefont {Jung}}, \bibinfo {author} {\bibfnamefont {L.}~\bibnamefont {Zhang}}, \bibinfo {author} {\bibfnamefont {M.}~\bibnamefont {Chen}}, \bibinfo {author} {\bibfnamefont {S.~J.}\ \bibnamefont {Parluhutan}}, \bibinfo {author} {\bibfnamefont {D.-H.}\ \bibnamefont {Kang}}, \bibinfo {author} {\bibfnamefont {C.}~\bibnamefont {Lee}}, \bibinfo {author} {\bibfnamefont {S.}~\bibnamefont {Assali}}, \emph {et~al.},\ }\href@noop {} {\bibfield  {journal} {\bibinfo  {journal} {Applied Physics Letters}\ }\textbf {\bibinfo {volume} {119}},\ \bibinfo {pages} {201101} (\bibinfo {year} {2021})}\BibitemShut {NoStop}%
\bibitem [{\citenamefont {Jung}\ \emph {et~al.}(2022)\citenamefont {Jung}, \citenamefont {Burt}, \citenamefont {Zhang}, \citenamefont {Kim}, \citenamefont {Joo}, \citenamefont {Chen}, \citenamefont {Assali}, \citenamefont {Moutanabbir}, \citenamefont {Tan},\ and\ \citenamefont {Nam}}]{jung2022optically}%
  \BibitemOpen
  \bibfield  {author} {\bibinfo {author} {\bibfnamefont {Y.}~\bibnamefont {Jung}}, \bibinfo {author} {\bibfnamefont {D.}~\bibnamefont {Burt}}, \bibinfo {author} {\bibfnamefont {L.}~\bibnamefont {Zhang}}, \bibinfo {author} {\bibfnamefont {Y.}~\bibnamefont {Kim}}, \bibinfo {author} {\bibfnamefont {H.-J.}\ \bibnamefont {Joo}}, \bibinfo {author} {\bibfnamefont {M.}~\bibnamefont {Chen}}, \bibinfo {author} {\bibfnamefont {S.}~\bibnamefont {Assali}}, \bibinfo {author} {\bibfnamefont {O.}~\bibnamefont {Moutanabbir}}, \bibinfo {author} {\bibfnamefont {C.~S.}\ \bibnamefont {Tan}},\ and\ \bibinfo {author} {\bibfnamefont {D.}~\bibnamefont {Nam}},\ }\href@noop {} {\bibfield  {journal} {\bibinfo  {journal} {Photonics Research}\ }\textbf {\bibinfo {volume} {10}},\ \bibinfo {pages} {1332} (\bibinfo {year} {2022})}\BibitemShut {NoStop}%
\bibitem [{\citenamefont {Li}\ \emph {et~al.}(2021)\citenamefont {Li}, \citenamefont {Peng}, \citenamefont {Liu}, \citenamefont {Zhou}, \citenamefont {Zheng}, \citenamefont {Xue}, \citenamefont {Zuo}, \citenamefont {Chen},\ and\ \citenamefont {Cheng}}]{li202130}%
  \BibitemOpen
  \bibfield  {author} {\bibinfo {author} {\bibfnamefont {X.}~\bibnamefont {Li}}, \bibinfo {author} {\bibfnamefont {L.}~\bibnamefont {Peng}}, \bibinfo {author} {\bibfnamefont {Z.}~\bibnamefont {Liu}}, \bibinfo {author} {\bibfnamefont {Z.}~\bibnamefont {Zhou}}, \bibinfo {author} {\bibfnamefont {J.}~\bibnamefont {Zheng}}, \bibinfo {author} {\bibfnamefont {C.}~\bibnamefont {Xue}}, \bibinfo {author} {\bibfnamefont {Y.}~\bibnamefont {Zuo}}, \bibinfo {author} {\bibfnamefont {B.}~\bibnamefont {Chen}},\ and\ \bibinfo {author} {\bibfnamefont {B.}~\bibnamefont {Cheng}},\ }\href@noop {} {\bibfield  {journal} {\bibinfo  {journal} {Photonics Research}\ }\textbf {\bibinfo {volume} {9}},\ \bibinfo {pages} {494} (\bibinfo {year} {2021})}\BibitemShut {NoStop}%
\bibitem [{\citenamefont {Liu}\ \emph {et~al.}(2022)\citenamefont {Liu}, \citenamefont {Zheng}, \citenamefont {Niu}, \citenamefont {Liu}, \citenamefont {Huang}, \citenamefont {Li}, \citenamefont {Zhang}, \citenamefont {Pang}, \citenamefont {Liu}, \citenamefont {Zuo} \emph {et~al.}}]{liu2022sn}%
  \BibitemOpen
  \bibfield  {author} {\bibinfo {author} {\bibfnamefont {X.}~\bibnamefont {Liu}}, \bibinfo {author} {\bibfnamefont {J.}~\bibnamefont {Zheng}}, \bibinfo {author} {\bibfnamefont {C.}~\bibnamefont {Niu}}, \bibinfo {author} {\bibfnamefont {T.}~\bibnamefont {Liu}}, \bibinfo {author} {\bibfnamefont {Q.}~\bibnamefont {Huang}}, \bibinfo {author} {\bibfnamefont {M.}~\bibnamefont {Li}}, \bibinfo {author} {\bibfnamefont {D.}~\bibnamefont {Zhang}}, \bibinfo {author} {\bibfnamefont {Y.}~\bibnamefont {Pang}}, \bibinfo {author} {\bibfnamefont {Z.}~\bibnamefont {Liu}}, \bibinfo {author} {\bibfnamefont {Y.}~\bibnamefont {Zuo}}, \emph {et~al.},\ }\href@noop {} {\bibfield  {journal} {\bibinfo  {journal} {Photonics Research}\ }\textbf {\bibinfo {volume} {10}},\ \bibinfo {pages} {1567} (\bibinfo {year} {2022})}\BibitemShut {NoStop}%
\bibitem [{\citenamefont {Luo}\ \emph {et~al.}(2022)\citenamefont {Luo}, \citenamefont {Assali}, \citenamefont {Atalla}, \citenamefont {Koelling}, \citenamefont {Attiaoui}, \citenamefont {Daligou}, \citenamefont {Mart{\'\i}}, \citenamefont {Arbiol},\ and\ \citenamefont {Moutanabbir}}]{luo2022extended}%
  \BibitemOpen
  \bibfield  {author} {\bibinfo {author} {\bibfnamefont {L.}~\bibnamefont {Luo}}, \bibinfo {author} {\bibfnamefont {S.}~\bibnamefont {Assali}}, \bibinfo {author} {\bibfnamefont {M.~R.}\ \bibnamefont {Atalla}}, \bibinfo {author} {\bibfnamefont {S.}~\bibnamefont {Koelling}}, \bibinfo {author} {\bibfnamefont {A.}~\bibnamefont {Attiaoui}}, \bibinfo {author} {\bibfnamefont {G.}~\bibnamefont {Daligou}}, \bibinfo {author} {\bibfnamefont {S.}~\bibnamefont {Mart{\'\i}}}, \bibinfo {author} {\bibfnamefont {J.}~\bibnamefont {Arbiol}},\ and\ \bibinfo {author} {\bibfnamefont {O.}~\bibnamefont {Moutanabbir}},\ }\href@noop {} {\bibfield  {journal} {\bibinfo  {journal} {ACS Photonics}\ }\textbf {\bibinfo {volume} {9}},\ \bibinfo {pages} {914} (\bibinfo {year} {2022})}\BibitemShut {NoStop}%
\bibitem [{\citenamefont {Marzban}\ \emph {et~al.}(2022)\citenamefont {Marzban}, \citenamefont {Seidel}, \citenamefont {Liu}, \citenamefont {Wu}, \citenamefont {Kiyek}, \citenamefont {Zoellner}, \citenamefont {Ikonic}, \citenamefont {Schulze}, \citenamefont {Gr{\"u}tzmacher}, \citenamefont {Capellini} \emph {et~al.}}]{marzban2022strain}%
  \BibitemOpen
  \bibfield  {author} {\bibinfo {author} {\bibfnamefont {B.}~\bibnamefont {Marzban}}, \bibinfo {author} {\bibfnamefont {L.}~\bibnamefont {Seidel}}, \bibinfo {author} {\bibfnamefont {T.}~\bibnamefont {Liu}}, \bibinfo {author} {\bibfnamefont {K.}~\bibnamefont {Wu}}, \bibinfo {author} {\bibfnamefont {V.}~\bibnamefont {Kiyek}}, \bibinfo {author} {\bibfnamefont {M.~H.}\ \bibnamefont {Zoellner}}, \bibinfo {author} {\bibfnamefont {Z.}~\bibnamefont {Ikonic}}, \bibinfo {author} {\bibfnamefont {J.}~\bibnamefont {Schulze}}, \bibinfo {author} {\bibfnamefont {D.}~\bibnamefont {Gr{\"u}tzmacher}}, \bibinfo {author} {\bibfnamefont {G.}~\bibnamefont {Capellini}}, \emph {et~al.},\ }\href@noop {} {\bibfield  {journal} {\bibinfo  {journal} {ACS Photonics}\ } (\bibinfo {year} {2022})}\BibitemShut {NoStop}%
\bibitem [{\citenamefont {Talamas~Simola}\ \emph {et~al.}(2021)\citenamefont {Talamas~Simola}, \citenamefont {Kiyek}, \citenamefont {Ballabio}, \citenamefont {Schlykow}, \citenamefont {Frigerio}, \citenamefont {Zucchetti}, \citenamefont {De~Iacovo}, \citenamefont {Colace}, \citenamefont {Yamamoto}, \citenamefont {Capellini} \emph {et~al.}}]{talamas2021cmos}%
  \BibitemOpen
  \bibfield  {author} {\bibinfo {author} {\bibfnamefont {E.}~\bibnamefont {Talamas~Simola}}, \bibinfo {author} {\bibfnamefont {V.}~\bibnamefont {Kiyek}}, \bibinfo {author} {\bibfnamefont {A.}~\bibnamefont {Ballabio}}, \bibinfo {author} {\bibfnamefont {V.}~\bibnamefont {Schlykow}}, \bibinfo {author} {\bibfnamefont {J.}~\bibnamefont {Frigerio}}, \bibinfo {author} {\bibfnamefont {C.}~\bibnamefont {Zucchetti}}, \bibinfo {author} {\bibfnamefont {A.}~\bibnamefont {De~Iacovo}}, \bibinfo {author} {\bibfnamefont {L.}~\bibnamefont {Colace}}, \bibinfo {author} {\bibfnamefont {Y.}~\bibnamefont {Yamamoto}}, \bibinfo {author} {\bibfnamefont {G.}~\bibnamefont {Capellini}}, \emph {et~al.},\ }\href@noop {} {\bibfield  {journal} {\bibinfo  {journal} {ACS photonics}\ }\textbf {\bibinfo {volume} {8}},\ \bibinfo {pages} {2166} (\bibinfo {year} {2021})}\BibitemShut {NoStop}%
\bibitem [{\citenamefont {Tran}\ \emph {et~al.}(2019)\citenamefont {Tran}, \citenamefont {Pham}, \citenamefont {Margetis}, \citenamefont {Zhou}, \citenamefont {Dou}, \citenamefont {Grant}, \citenamefont {Grant}, \citenamefont {Al-Kabi}, \citenamefont {Sun}, \citenamefont {Soref} \emph {et~al.}}]{tran2019si}%
  \BibitemOpen
  \bibfield  {author} {\bibinfo {author} {\bibfnamefont {H.}~\bibnamefont {Tran}}, \bibinfo {author} {\bibfnamefont {T.}~\bibnamefont {Pham}}, \bibinfo {author} {\bibfnamefont {J.}~\bibnamefont {Margetis}}, \bibinfo {author} {\bibfnamefont {Y.}~\bibnamefont {Zhou}}, \bibinfo {author} {\bibfnamefont {W.}~\bibnamefont {Dou}}, \bibinfo {author} {\bibfnamefont {P.~C.}\ \bibnamefont {Grant}}, \bibinfo {author} {\bibfnamefont {J.~M.}\ \bibnamefont {Grant}}, \bibinfo {author} {\bibfnamefont {S.}~\bibnamefont {Al-Kabi}}, \bibinfo {author} {\bibfnamefont {G.}~\bibnamefont {Sun}}, \bibinfo {author} {\bibfnamefont {R.~A.}\ \bibnamefont {Soref}}, \emph {et~al.},\ }\href@noop {} {\bibfield  {journal} {\bibinfo  {journal} {ACS Photonics}\ }\textbf {\bibinfo {volume} {6}},\ \bibinfo {pages} {2807} (\bibinfo {year} {2019})}\BibitemShut {NoStop}%
\bibitem [{\citenamefont {Xu}\ \emph {et~al.}(2019)\citenamefont {Xu}, \citenamefont {Wang}, \citenamefont {Huang}, \citenamefont {Dong}, \citenamefont {Masudy-Panah}, \citenamefont {Wang}, \citenamefont {Gong},\ and\ \citenamefont {Yeo}}]{xu2019high}%
  \BibitemOpen
  \bibfield  {author} {\bibinfo {author} {\bibfnamefont {S.}~\bibnamefont {Xu}}, \bibinfo {author} {\bibfnamefont {W.}~\bibnamefont {Wang}}, \bibinfo {author} {\bibfnamefont {Y.-C.}\ \bibnamefont {Huang}}, \bibinfo {author} {\bibfnamefont {Y.}~\bibnamefont {Dong}}, \bibinfo {author} {\bibfnamefont {S.}~\bibnamefont {Masudy-Panah}}, \bibinfo {author} {\bibfnamefont {H.}~\bibnamefont {Wang}}, \bibinfo {author} {\bibfnamefont {X.}~\bibnamefont {Gong}},\ and\ \bibinfo {author} {\bibfnamefont {Y.-C.}\ \bibnamefont {Yeo}},\ }\href@noop {} {\bibfield  {journal} {\bibinfo  {journal} {Optics express}\ }\textbf {\bibinfo {volume} {27}},\ \bibinfo {pages} {5798} (\bibinfo {year} {2019})}\BibitemShut {NoStop}%
\bibitem [{\citenamefont {Zhou}\ \emph {et~al.}(2020)\citenamefont {Zhou}, \citenamefont {Miao}, \citenamefont {Ojo}, \citenamefont {Tran}, \citenamefont {Abernathy}, \citenamefont {Grant}, \citenamefont {Amoah}, \citenamefont {Salamo}, \citenamefont {Du}, \citenamefont {Liu} \emph {et~al.}}]{zhou2020electrically}%
  \BibitemOpen
  \bibfield  {author} {\bibinfo {author} {\bibfnamefont {Y.}~\bibnamefont {Zhou}}, \bibinfo {author} {\bibfnamefont {Y.}~\bibnamefont {Miao}}, \bibinfo {author} {\bibfnamefont {S.}~\bibnamefont {Ojo}}, \bibinfo {author} {\bibfnamefont {H.}~\bibnamefont {Tran}}, \bibinfo {author} {\bibfnamefont {G.}~\bibnamefont {Abernathy}}, \bibinfo {author} {\bibfnamefont {J.~M.}\ \bibnamefont {Grant}}, \bibinfo {author} {\bibfnamefont {S.}~\bibnamefont {Amoah}}, \bibinfo {author} {\bibfnamefont {G.}~\bibnamefont {Salamo}}, \bibinfo {author} {\bibfnamefont {W.}~\bibnamefont {Du}}, \bibinfo {author} {\bibfnamefont {J.}~\bibnamefont {Liu}}, \emph {et~al.},\ }\href@noop {} {\bibfield  {journal} {\bibinfo  {journal} {Optica}\ }\textbf {\bibinfo {volume} {7}},\ \bibinfo {pages} {924} (\bibinfo {year} {2020})}\BibitemShut {NoStop}%
\bibitem [{\citenamefont {Daligou}\ \emph {et~al.}(2023)\citenamefont {Daligou}, \citenamefont {Soref}, \citenamefont {Attiaoui}, \citenamefont {Hossain}, \citenamefont {Atalla}, \citenamefont {Del~Vecchio},\ and\ \citenamefont {Moutanabbir}}]{daligou2023group}%
  \BibitemOpen
  \bibfield  {author} {\bibinfo {author} {\bibfnamefont {G.}~\bibnamefont {Daligou}}, \bibinfo {author} {\bibfnamefont {R.}~\bibnamefont {Soref}}, \bibinfo {author} {\bibfnamefont {A.}~\bibnamefont {Attiaoui}}, \bibinfo {author} {\bibfnamefont {J.}~\bibnamefont {Hossain}}, \bibinfo {author} {\bibfnamefont {M.~R.}\ \bibnamefont {Atalla}}, \bibinfo {author} {\bibfnamefont {P.}~\bibnamefont {Del~Vecchio}},\ and\ \bibinfo {author} {\bibfnamefont {O.}~\bibnamefont {Moutanabbir}},\ }\href@noop {} {\bibfield  {journal} {\bibinfo  {journal} {IEEE Journal of Photovoltaics}\ } (\bibinfo {year} {2023})}\BibitemShut {NoStop}%
\bibitem [{\citenamefont {Assali}\ \emph {et~al.}(2019)\citenamefont {Assali}, \citenamefont {Albani}, \citenamefont {Bergamaschini}, \citenamefont {Verheijen}, \citenamefont {Li}, \citenamefont {K{\"o}lling}, \citenamefont {Gagliano}, \citenamefont {Bakkers},\ and\ \citenamefont {Miglio}}]{assali2019strain}%
  \BibitemOpen
  \bibfield  {author} {\bibinfo {author} {\bibfnamefont {S.}~\bibnamefont {Assali}}, \bibinfo {author} {\bibfnamefont {M.}~\bibnamefont {Albani}}, \bibinfo {author} {\bibfnamefont {R.}~\bibnamefont {Bergamaschini}}, \bibinfo {author} {\bibfnamefont {M.~A.}\ \bibnamefont {Verheijen}}, \bibinfo {author} {\bibfnamefont {A.}~\bibnamefont {Li}}, \bibinfo {author} {\bibfnamefont {S.}~\bibnamefont {K{\"o}lling}}, \bibinfo {author} {\bibfnamefont {L.}~\bibnamefont {Gagliano}}, \bibinfo {author} {\bibfnamefont {E.~P.}\ \bibnamefont {Bakkers}},\ and\ \bibinfo {author} {\bibfnamefont {L.}~\bibnamefont {Miglio}},\ }\href@noop {} {\bibfield  {journal} {\bibinfo  {journal} {Applied Physics Letters}\ }\textbf {\bibinfo {volume} {115}},\ \bibinfo {pages} {113102} (\bibinfo {year} {2019})}\BibitemShut {NoStop}%
\bibitem [{\citenamefont {Aubin}\ \emph {et~al.}(2017)\citenamefont {Aubin}, \citenamefont {Hartmann}, \citenamefont {Gassenq}, \citenamefont {Rouviere}, \citenamefont {Robin}, \citenamefont {Delaye}, \citenamefont {Cooper}, \citenamefont {Mollard}, \citenamefont {Reboud},\ and\ \citenamefont {Calvo}}]{aubin2017growth}%
  \BibitemOpen
  \bibfield  {author} {\bibinfo {author} {\bibfnamefont {J.}~\bibnamefont {Aubin}}, \bibinfo {author} {\bibfnamefont {J.}~\bibnamefont {Hartmann}}, \bibinfo {author} {\bibfnamefont {A.}~\bibnamefont {Gassenq}}, \bibinfo {author} {\bibfnamefont {J.}~\bibnamefont {Rouviere}}, \bibinfo {author} {\bibfnamefont {E.}~\bibnamefont {Robin}}, \bibinfo {author} {\bibfnamefont {V.}~\bibnamefont {Delaye}}, \bibinfo {author} {\bibfnamefont {D.}~\bibnamefont {Cooper}}, \bibinfo {author} {\bibfnamefont {N.}~\bibnamefont {Mollard}}, \bibinfo {author} {\bibfnamefont {V.}~\bibnamefont {Reboud}},\ and\ \bibinfo {author} {\bibfnamefont {V.}~\bibnamefont {Calvo}},\ }\href@noop {} {\bibfield  {journal} {\bibinfo  {journal} {Semiconductor Science and Technology}\ }\textbf {\bibinfo {volume} {32}},\ \bibinfo {pages} {094006} (\bibinfo {year} {2017})}\BibitemShut {NoStop}%
\bibitem [{\citenamefont {Dou}\ \emph {et~al.}(2018)\citenamefont {Dou}, \citenamefont {Benamara}, \citenamefont {Mosleh}, \citenamefont {Margetis}, \citenamefont {Grant}, \citenamefont {Zhou}, \citenamefont {Al-Kabi}, \citenamefont {Du}, \citenamefont {Tolle}, \citenamefont {Li} \emph {et~al.}}]{dou2018investigation}%
  \BibitemOpen
  \bibfield  {author} {\bibinfo {author} {\bibfnamefont {W.}~\bibnamefont {Dou}}, \bibinfo {author} {\bibfnamefont {M.}~\bibnamefont {Benamara}}, \bibinfo {author} {\bibfnamefont {A.}~\bibnamefont {Mosleh}}, \bibinfo {author} {\bibfnamefont {J.}~\bibnamefont {Margetis}}, \bibinfo {author} {\bibfnamefont {P.}~\bibnamefont {Grant}}, \bibinfo {author} {\bibfnamefont {Y.}~\bibnamefont {Zhou}}, \bibinfo {author} {\bibfnamefont {S.}~\bibnamefont {Al-Kabi}}, \bibinfo {author} {\bibfnamefont {W.}~\bibnamefont {Du}}, \bibinfo {author} {\bibfnamefont {J.}~\bibnamefont {Tolle}}, \bibinfo {author} {\bibfnamefont {B.}~\bibnamefont {Li}}, \emph {et~al.},\ }\href@noop {} {\bibfield  {journal} {\bibinfo  {journal} {Scientific reports}\ }\textbf {\bibinfo {volume} {8}},\ \bibinfo {pages} {5640} (\bibinfo {year} {2018})}\BibitemShut {NoStop}%
\bibitem [{\citenamefont {Assali}\ \emph {et~al.}(2018)\citenamefont {Assali}, \citenamefont {Nicolas}, \citenamefont {Mukherjee}, \citenamefont {Dijkstra},\ and\ \citenamefont {Moutanabbir}}]{assali2018atomically}%
  \BibitemOpen
  \bibfield  {author} {\bibinfo {author} {\bibfnamefont {S.}~\bibnamefont {Assali}}, \bibinfo {author} {\bibfnamefont {J.}~\bibnamefont {Nicolas}}, \bibinfo {author} {\bibfnamefont {S.}~\bibnamefont {Mukherjee}}, \bibinfo {author} {\bibfnamefont {A.}~\bibnamefont {Dijkstra}},\ and\ \bibinfo {author} {\bibfnamefont {O.}~\bibnamefont {Moutanabbir}},\ }\href@noop {} {\bibfield  {journal} {\bibinfo  {journal} {Applied Physics Letters}\ }\textbf {\bibinfo {volume} {112}},\ \bibinfo {pages} {251903} (\bibinfo {year} {2018})}\BibitemShut {NoStop}%
\bibitem [{\citenamefont {Atalla}\ \emph {et~al.}(2023{\natexlab{b}})\citenamefont {Atalla}, \citenamefont {Assali}, \citenamefont {Koelling}, \citenamefont {Attiaoui},\ and\ \citenamefont {Moutanabbir}}]{atalla2023dark}%
  \BibitemOpen
  \bibfield  {author} {\bibinfo {author} {\bibfnamefont {M.~R.}\ \bibnamefont {Atalla}}, \bibinfo {author} {\bibfnamefont {S.}~\bibnamefont {Assali}}, \bibinfo {author} {\bibfnamefont {S.}~\bibnamefont {Koelling}}, \bibinfo {author} {\bibfnamefont {A.}~\bibnamefont {Attiaoui}},\ and\ \bibinfo {author} {\bibfnamefont {O.}~\bibnamefont {Moutanabbir}},\ }\href@noop {} {\bibfield  {journal} {\bibinfo  {journal} {Applied Physics Letters}\ }\textbf {\bibinfo {volume} {122}} (\bibinfo {year} {2023}{\natexlab{b}})}\BibitemShut {NoStop}%
\bibitem [{\citenamefont {Elbaz}\ \emph {et~al.}(2020{\natexlab{b}})\citenamefont {Elbaz}, \citenamefont {Arefin}, \citenamefont {Sakat}, \citenamefont {Wang}, \citenamefont {Herth}, \citenamefont {Patriarche}, \citenamefont {Foti}, \citenamefont {Ossikovski}, \citenamefont {Sauvage}, \citenamefont {Checoury} \emph {et~al.}}]{elbaz2020reduced}%
  \BibitemOpen
  \bibfield  {author} {\bibinfo {author} {\bibfnamefont {A.}~\bibnamefont {Elbaz}}, \bibinfo {author} {\bibfnamefont {R.}~\bibnamefont {Arefin}}, \bibinfo {author} {\bibfnamefont {E.}~\bibnamefont {Sakat}}, \bibinfo {author} {\bibfnamefont {B.}~\bibnamefont {Wang}}, \bibinfo {author} {\bibfnamefont {E.}~\bibnamefont {Herth}}, \bibinfo {author} {\bibfnamefont {G.}~\bibnamefont {Patriarche}}, \bibinfo {author} {\bibfnamefont {A.}~\bibnamefont {Foti}}, \bibinfo {author} {\bibfnamefont {R.}~\bibnamefont {Ossikovski}}, \bibinfo {author} {\bibfnamefont {S.}~\bibnamefont {Sauvage}}, \bibinfo {author} {\bibfnamefont {X.}~\bibnamefont {Checoury}}, \emph {et~al.},\ }\href@noop {} {\bibfield  {journal} {\bibinfo  {journal} {ACS photonics}\ }\textbf {\bibinfo {volume} {7}},\ \bibinfo {pages} {2713} (\bibinfo {year} {2020}{\natexlab{b}})}\BibitemShut {NoStop}%
\bibitem [{\citenamefont {Thai}\ \emph {et~al.}(2018)\citenamefont {Thai}, \citenamefont {Pauc}, \citenamefont {Aubin}, \citenamefont {Bertrand}, \citenamefont {Chr{\'e}tien}, \citenamefont {Delaye}, \citenamefont {Chelnokov}, \citenamefont {Hartmann}, \citenamefont {Reboud},\ and\ \citenamefont {Calvo}}]{thai2018gesn}%
  \BibitemOpen
  \bibfield  {author} {\bibinfo {author} {\bibfnamefont {Q.~M.}\ \bibnamefont {Thai}}, \bibinfo {author} {\bibfnamefont {N.}~\bibnamefont {Pauc}}, \bibinfo {author} {\bibfnamefont {J.}~\bibnamefont {Aubin}}, \bibinfo {author} {\bibfnamefont {M.}~\bibnamefont {Bertrand}}, \bibinfo {author} {\bibfnamefont {J.}~\bibnamefont {Chr{\'e}tien}}, \bibinfo {author} {\bibfnamefont {V.}~\bibnamefont {Delaye}}, \bibinfo {author} {\bibfnamefont {A.}~\bibnamefont {Chelnokov}}, \bibinfo {author} {\bibfnamefont {J.-M.}\ \bibnamefont {Hartmann}}, \bibinfo {author} {\bibfnamefont {V.}~\bibnamefont {Reboud}},\ and\ \bibinfo {author} {\bibfnamefont {V.}~\bibnamefont {Calvo}},\ }\href@noop {} {\bibfield  {journal} {\bibinfo  {journal} {Optics express}\ }\textbf {\bibinfo {volume} {26}},\ \bibinfo {pages} {32500} (\bibinfo {year} {2018})}\BibitemShut {NoStop}%
\bibitem [{\citenamefont {Kim}\ \emph {et~al.}(2022)\citenamefont {Kim}, \citenamefont {Assali}, \citenamefont {Burt}, \citenamefont {Jung}, \citenamefont {Joo}, \citenamefont {Chen}, \citenamefont {Ikonic}, \citenamefont {Moutanabbir},\ and\ \citenamefont {Nam}}]{kim2022enhanced}%
  \BibitemOpen
  \bibfield  {author} {\bibinfo {author} {\bibfnamefont {Y.}~\bibnamefont {Kim}}, \bibinfo {author} {\bibfnamefont {S.}~\bibnamefont {Assali}}, \bibinfo {author} {\bibfnamefont {D.}~\bibnamefont {Burt}}, \bibinfo {author} {\bibfnamefont {Y.}~\bibnamefont {Jung}}, \bibinfo {author} {\bibfnamefont {H.-J.}\ \bibnamefont {Joo}}, \bibinfo {author} {\bibfnamefont {M.}~\bibnamefont {Chen}}, \bibinfo {author} {\bibfnamefont {Z.}~\bibnamefont {Ikonic}}, \bibinfo {author} {\bibfnamefont {O.}~\bibnamefont {Moutanabbir}},\ and\ \bibinfo {author} {\bibfnamefont {D.}~\bibnamefont {Nam}},\ }\href@noop {} {\bibfield  {journal} {\bibinfo  {journal} {Advanced Optical Materials}\ }\textbf {\bibinfo {volume} {10}},\ \bibinfo {pages} {2101213} (\bibinfo {year} {2022})}\BibitemShut {NoStop}%
\bibitem [{\citenamefont {Stange}\ \emph {et~al.}(2016)\citenamefont {Stange}, \citenamefont {Wirths}, \citenamefont {Geiger}, \citenamefont {Schulte-Braucks}, \citenamefont {Marzban}, \citenamefont {Von Den~Driesch}, \citenamefont {Mussler}, \citenamefont {Zabel}, \citenamefont {Stoica}, \citenamefont {Hartmann} \emph {et~al.}}]{stange2016optically}%
  \BibitemOpen
  \bibfield  {author} {\bibinfo {author} {\bibfnamefont {D.}~\bibnamefont {Stange}}, \bibinfo {author} {\bibfnamefont {S.}~\bibnamefont {Wirths}}, \bibinfo {author} {\bibfnamefont {R.}~\bibnamefont {Geiger}}, \bibinfo {author} {\bibfnamefont {C.}~\bibnamefont {Schulte-Braucks}}, \bibinfo {author} {\bibfnamefont {B.}~\bibnamefont {Marzban}}, \bibinfo {author} {\bibfnamefont {N.}~\bibnamefont {Von Den~Driesch}}, \bibinfo {author} {\bibfnamefont {G.}~\bibnamefont {Mussler}}, \bibinfo {author} {\bibfnamefont {T.}~\bibnamefont {Zabel}}, \bibinfo {author} {\bibfnamefont {T.}~\bibnamefont {Stoica}}, \bibinfo {author} {\bibfnamefont {J.-M.}\ \bibnamefont {Hartmann}}, \emph {et~al.},\ }\href@noop {} {\bibfield  {journal} {\bibinfo  {journal} {ACS photonics}\ }\textbf {\bibinfo {volume} {3}},\ \bibinfo {pages} {1279} (\bibinfo {year} {2016})}\BibitemShut {NoStop}%
\bibitem [{\citenamefont {Burt}\ \emph {et~al.}(2021)\citenamefont {Burt}, \citenamefont {Joo}, \citenamefont {Jung}, \citenamefont {Kim}, \citenamefont {Chen}, \citenamefont {Huang},\ and\ \citenamefont {Nam}}]{burt2021strain}%
  \BibitemOpen
  \bibfield  {author} {\bibinfo {author} {\bibfnamefont {D.}~\bibnamefont {Burt}}, \bibinfo {author} {\bibfnamefont {H.-J.}\ \bibnamefont {Joo}}, \bibinfo {author} {\bibfnamefont {Y.}~\bibnamefont {Jung}}, \bibinfo {author} {\bibfnamefont {Y.}~\bibnamefont {Kim}}, \bibinfo {author} {\bibfnamefont {M.}~\bibnamefont {Chen}}, \bibinfo {author} {\bibfnamefont {Y.-C.}\ \bibnamefont {Huang}},\ and\ \bibinfo {author} {\bibfnamefont {D.}~\bibnamefont {Nam}},\ }\href@noop {} {\bibfield  {journal} {\bibinfo  {journal} {Optics Express}\ }\textbf {\bibinfo {volume} {29}},\ \bibinfo {pages} {28959} (\bibinfo {year} {2021})}\BibitemShut {NoStop}%
\bibitem [{\citenamefont {Galluccio}\ \emph {et~al.}(2020)\citenamefont {Galluccio}, \citenamefont {Doherty}, \citenamefont {Biswas}, \citenamefont {Holmes},\ and\ \citenamefont {Duffy}}]{galluccio2020field}%
  \BibitemOpen
  \bibfield  {author} {\bibinfo {author} {\bibfnamefont {E.}~\bibnamefont {Galluccio}}, \bibinfo {author} {\bibfnamefont {J.}~\bibnamefont {Doherty}}, \bibinfo {author} {\bibfnamefont {S.}~\bibnamefont {Biswas}}, \bibinfo {author} {\bibfnamefont {J.~D.}\ \bibnamefont {Holmes}},\ and\ \bibinfo {author} {\bibfnamefont {R.}~\bibnamefont {Duffy}},\ }\href@noop {} {\bibfield  {journal} {\bibinfo  {journal} {ACS Applied Electronic Materials}\ }\textbf {\bibinfo {volume} {2}},\ \bibinfo {pages} {1226} (\bibinfo {year} {2020})}\BibitemShut {NoStop}%
\bibitem [{\citenamefont {Sistani}\ \emph {et~al.}(2018)\citenamefont {Sistani}, \citenamefont {Seifner}, \citenamefont {Bartmann}, \citenamefont {Smoliner}, \citenamefont {Lugstein},\ and\ \citenamefont {Barth}}]{sistani2018electrical}%
  \BibitemOpen
  \bibfield  {author} {\bibinfo {author} {\bibfnamefont {M.}~\bibnamefont {Sistani}}, \bibinfo {author} {\bibfnamefont {M.}~\bibnamefont {Seifner}}, \bibinfo {author} {\bibfnamefont {M.}~\bibnamefont {Bartmann}}, \bibinfo {author} {\bibfnamefont {J.}~\bibnamefont {Smoliner}}, \bibinfo {author} {\bibfnamefont {A.}~\bibnamefont {Lugstein}},\ and\ \bibinfo {author} {\bibfnamefont {S.}~\bibnamefont {Barth}},\ }\href@noop {} {\bibfield  {journal} {\bibinfo  {journal} {Nanoscale}\ }\textbf {\bibinfo {volume} {10}},\ \bibinfo {pages} {19443} (\bibinfo {year} {2018})}\BibitemShut {NoStop}%
\bibitem [{\citenamefont {Seifner}\ \emph {et~al.}(2019)\citenamefont {Seifner}, \citenamefont {Dijkstra}, \citenamefont {Bernardi}, \citenamefont {Steiger-Thirsfeld}, \citenamefont {Sistani}, \citenamefont {Lugstein}, \citenamefont {Haverkort},\ and\ \citenamefont {Barth}}]{seifner2019epitaxial}%
  \BibitemOpen
  \bibfield  {author} {\bibinfo {author} {\bibfnamefont {M.~S.}\ \bibnamefont {Seifner}}, \bibinfo {author} {\bibfnamefont {A.}~\bibnamefont {Dijkstra}}, \bibinfo {author} {\bibfnamefont {J.}~\bibnamefont {Bernardi}}, \bibinfo {author} {\bibfnamefont {A.}~\bibnamefont {Steiger-Thirsfeld}}, \bibinfo {author} {\bibfnamefont {M.}~\bibnamefont {Sistani}}, \bibinfo {author} {\bibfnamefont {A.}~\bibnamefont {Lugstein}}, \bibinfo {author} {\bibfnamefont {J.~E.}\ \bibnamefont {Haverkort}},\ and\ \bibinfo {author} {\bibfnamefont {S.}~\bibnamefont {Barth}},\ }\href@noop {} {\bibfield  {journal} {\bibinfo  {journal} {ACS nano}\ }\textbf {\bibinfo {volume} {13}},\ \bibinfo {pages} {8047} (\bibinfo {year} {2019})}\BibitemShut {NoStop}%
\bibitem [{\citenamefont {Meng}\ \emph {et~al.}(2016)\citenamefont {Meng}, \citenamefont {Fenrich}, \citenamefont {Braun}, \citenamefont {McVittie}, \citenamefont {Marshall}, \citenamefont {Harris},\ and\ \citenamefont {McIntyre}}]{meng2016core}%
  \BibitemOpen
  \bibfield  {author} {\bibinfo {author} {\bibfnamefont {A.~C.}\ \bibnamefont {Meng}}, \bibinfo {author} {\bibfnamefont {C.~S.}\ \bibnamefont {Fenrich}}, \bibinfo {author} {\bibfnamefont {M.~R.}\ \bibnamefont {Braun}}, \bibinfo {author} {\bibfnamefont {J.~P.}\ \bibnamefont {McVittie}}, \bibinfo {author} {\bibfnamefont {A.~F.}\ \bibnamefont {Marshall}}, \bibinfo {author} {\bibfnamefont {J.~S.}\ \bibnamefont {Harris}},\ and\ \bibinfo {author} {\bibfnamefont {P.~C.}\ \bibnamefont {McIntyre}},\ }\href@noop {} {\bibfield  {journal} {\bibinfo  {journal} {Nano letters}\ }\textbf {\bibinfo {volume} {16}},\ \bibinfo {pages} {7521} (\bibinfo {year} {2016})}\BibitemShut {NoStop}%
\bibitem [{\citenamefont {Assali}\ \emph {et~al.}(2017)\citenamefont {Assali}, \citenamefont {Dijkstra}, \citenamefont {Li}, \citenamefont {Koelling}, \citenamefont {Verheijen}, \citenamefont {Gagliano}, \citenamefont {Von Den~Driesch}, \citenamefont {Buca}, \citenamefont {Koenraad}, \citenamefont {Haverkort} \emph {et~al.}}]{assali2017growth}%
  \BibitemOpen
  \bibfield  {author} {\bibinfo {author} {\bibfnamefont {S.}~\bibnamefont {Assali}}, \bibinfo {author} {\bibfnamefont {A.}~\bibnamefont {Dijkstra}}, \bibinfo {author} {\bibfnamefont {A.}~\bibnamefont {Li}}, \bibinfo {author} {\bibfnamefont {S.}~\bibnamefont {Koelling}}, \bibinfo {author} {\bibfnamefont {M.}~\bibnamefont {Verheijen}}, \bibinfo {author} {\bibfnamefont {L.}~\bibnamefont {Gagliano}}, \bibinfo {author} {\bibfnamefont {N.}~\bibnamefont {Von Den~Driesch}}, \bibinfo {author} {\bibfnamefont {D.}~\bibnamefont {Buca}}, \bibinfo {author} {\bibfnamefont {P.}~\bibnamefont {Koenraad}}, \bibinfo {author} {\bibfnamefont {J.}~\bibnamefont {Haverkort}}, \emph {et~al.},\ }\href@noop {} {\bibfield  {journal} {\bibinfo  {journal} {Nano letters}\ }\textbf {\bibinfo {volume} {17}},\ \bibinfo {pages} {1538} (\bibinfo {year} {2017})}\BibitemShut {NoStop}%
\bibitem [{\citenamefont {Lentz}\ \emph {et~al.}(2023)\citenamefont {Lentz}, \citenamefont {Woicik}, \citenamefont {Bergschneider}, \citenamefont {Davis}, \citenamefont {Mehta}, \citenamefont {Cho},\ and\ \citenamefont {McIntyre}}]{lentz2023local}%
  \BibitemOpen
  \bibfield  {author} {\bibinfo {author} {\bibfnamefont {J.~Z.}\ \bibnamefont {Lentz}}, \bibinfo {author} {\bibfnamefont {J.}~\bibnamefont {Woicik}}, \bibinfo {author} {\bibfnamefont {M.}~\bibnamefont {Bergschneider}}, \bibinfo {author} {\bibfnamefont {R.}~\bibnamefont {Davis}}, \bibinfo {author} {\bibfnamefont {A.}~\bibnamefont {Mehta}}, \bibinfo {author} {\bibfnamefont {K.}~\bibnamefont {Cho}},\ and\ \bibinfo {author} {\bibfnamefont {P.~C.}\ \bibnamefont {McIntyre}},\ }\href@noop {} {\bibfield  {journal} {\bibinfo  {journal} {Applied Physics Letters}\ }\textbf {\bibinfo {volume} {122}} (\bibinfo {year} {2023})}\BibitemShut {NoStop}%
\bibitem [{\citenamefont {Assali}\ \emph {et~al.}(2020)\citenamefont {Assali}, \citenamefont {Bergamaschini}, \citenamefont {Scalise}, \citenamefont {Verheijen}, \citenamefont {Albani}, \citenamefont {Dijkstra}, \citenamefont {Li}, \citenamefont {Koelling}, \citenamefont {Bakkers}, \citenamefont {Montalenti} \emph {et~al.}}]{assali2020kinetic}%
  \BibitemOpen
  \bibfield  {author} {\bibinfo {author} {\bibfnamefont {S.}~\bibnamefont {Assali}}, \bibinfo {author} {\bibfnamefont {R.}~\bibnamefont {Bergamaschini}}, \bibinfo {author} {\bibfnamefont {E.}~\bibnamefont {Scalise}}, \bibinfo {author} {\bibfnamefont {M.~A.}\ \bibnamefont {Verheijen}}, \bibinfo {author} {\bibfnamefont {M.}~\bibnamefont {Albani}}, \bibinfo {author} {\bibfnamefont {A.}~\bibnamefont {Dijkstra}}, \bibinfo {author} {\bibfnamefont {A.}~\bibnamefont {Li}}, \bibinfo {author} {\bibfnamefont {S.}~\bibnamefont {Koelling}}, \bibinfo {author} {\bibfnamefont {E.~P.}\ \bibnamefont {Bakkers}}, \bibinfo {author} {\bibfnamefont {F.}~\bibnamefont {Montalenti}}, \emph {et~al.},\ }\href@noop {} {\bibfield  {journal} {\bibinfo  {journal} {ACS nano}\ }\textbf {\bibinfo {volume} {14}},\ \bibinfo {pages} {2445} (\bibinfo {year} {2020})}\BibitemShut {NoStop}%
\bibitem [{\citenamefont {Royo}\ \emph {et~al.}(2017)\citenamefont {Royo}, \citenamefont {De~Luca}, \citenamefont {Rurali},\ and\ \citenamefont {Zardo}}]{royo2017review}%
  \BibitemOpen
  \bibfield  {author} {\bibinfo {author} {\bibfnamefont {M.}~\bibnamefont {Royo}}, \bibinfo {author} {\bibfnamefont {M.}~\bibnamefont {De~Luca}}, \bibinfo {author} {\bibfnamefont {R.}~\bibnamefont {Rurali}},\ and\ \bibinfo {author} {\bibfnamefont {I.}~\bibnamefont {Zardo}},\ }\href@noop {} {\bibfield  {journal} {\bibinfo  {journal} {Journal of Physics D: Applied Physics}\ }\textbf {\bibinfo {volume} {50}},\ \bibinfo {pages} {143001} (\bibinfo {year} {2017})}\BibitemShut {NoStop}%
\bibitem [{\citenamefont {Albani}\ \emph {et~al.}(2018)\citenamefont {Albani}, \citenamefont {Assali}, \citenamefont {Verheijen}, \citenamefont {Koelling}, \citenamefont {Bergamaschini}, \citenamefont {Pezzoli}, \citenamefont {Bakkers},\ and\ \citenamefont {Miglio}}]{albani2018critical}%
  \BibitemOpen
  \bibfield  {author} {\bibinfo {author} {\bibfnamefont {M.}~\bibnamefont {Albani}}, \bibinfo {author} {\bibfnamefont {S.}~\bibnamefont {Assali}}, \bibinfo {author} {\bibfnamefont {M.~A.}\ \bibnamefont {Verheijen}}, \bibinfo {author} {\bibfnamefont {S.}~\bibnamefont {Koelling}}, \bibinfo {author} {\bibfnamefont {R.}~\bibnamefont {Bergamaschini}}, \bibinfo {author} {\bibfnamefont {F.}~\bibnamefont {Pezzoli}}, \bibinfo {author} {\bibfnamefont {E.~P.}\ \bibnamefont {Bakkers}},\ and\ \bibinfo {author} {\bibfnamefont {L.}~\bibnamefont {Miglio}},\ }\href@noop {} {\bibfield  {journal} {\bibinfo  {journal} {Nanoscale}\ }\textbf {\bibinfo {volume} {10}},\ \bibinfo {pages} {7250} (\bibinfo {year} {2018})}\BibitemShut {NoStop}%
\bibitem [{\citenamefont {Kim}\ \emph {et~al.}(2023)\citenamefont {Kim}, \citenamefont {Assali}, \citenamefont {Joo}, \citenamefont {Koelling}, \citenamefont {Chen}, \citenamefont {Luo}, \citenamefont {Shi}, \citenamefont {Burt}, \citenamefont {Ikonic}, \citenamefont {Nam} \emph {et~al.}}]{kim2023short}%
  \BibitemOpen
  \bibfield  {author} {\bibinfo {author} {\bibfnamefont {Y.}~\bibnamefont {Kim}}, \bibinfo {author} {\bibfnamefont {S.}~\bibnamefont {Assali}}, \bibinfo {author} {\bibfnamefont {H.-J.}\ \bibnamefont {Joo}}, \bibinfo {author} {\bibfnamefont {S.}~\bibnamefont {Koelling}}, \bibinfo {author} {\bibfnamefont {M.}~\bibnamefont {Chen}}, \bibinfo {author} {\bibfnamefont {L.}~\bibnamefont {Luo}}, \bibinfo {author} {\bibfnamefont {X.}~\bibnamefont {Shi}}, \bibinfo {author} {\bibfnamefont {D.}~\bibnamefont {Burt}}, \bibinfo {author} {\bibfnamefont {Z.}~\bibnamefont {Ikonic}}, \bibinfo {author} {\bibfnamefont {D.}~\bibnamefont {Nam}}, \emph {et~al.},\ }\href@noop {} {\bibfield  {journal} {\bibinfo  {journal} {Nature Communications}\ }\textbf {\bibinfo {volume} {14}},\ \bibinfo {pages} {4393} (\bibinfo {year} {2023})}\BibitemShut {NoStop}%
\bibitem [{\citenamefont {Bouthillier}\ \emph {et~al.}(2020)\citenamefont {Bouthillier}, \citenamefont {Assali}, \citenamefont {Nicolas},\ and\ \citenamefont {Moutanabbir}}]{bouthillier2020decoupling}%
  \BibitemOpen
  \bibfield  {author} {\bibinfo {author} {\bibfnamefont {{\'E}.}~\bibnamefont {Bouthillier}}, \bibinfo {author} {\bibfnamefont {S.}~\bibnamefont {Assali}}, \bibinfo {author} {\bibfnamefont {J.}~\bibnamefont {Nicolas}},\ and\ \bibinfo {author} {\bibfnamefont {O.}~\bibnamefont {Moutanabbir}},\ }\href@noop {} {\bibfield  {journal} {\bibinfo  {journal} {Semiconductor Science and Technology}\ }\textbf {\bibinfo {volume} {35}},\ \bibinfo {pages} {095006} (\bibinfo {year} {2020})}\BibitemShut {NoStop}%
\bibitem [{\citenamefont {Nelin}\ and\ \citenamefont {Nilsson}(1972)}]{nelin1972phonon}%
  \BibitemOpen
  \bibfield  {author} {\bibinfo {author} {\bibfnamefont {G.}~\bibnamefont {Nelin}}\ and\ \bibinfo {author} {\bibfnamefont {G.}~\bibnamefont {Nilsson}},\ }\href@noop {} {\bibfield  {journal} {\bibinfo  {journal} {Physical Review B}\ }\textbf {\bibinfo {volume} {5}},\ \bibinfo {pages} {3151} (\bibinfo {year} {1972})}\BibitemShut {NoStop}%
\bibitem [{\citenamefont {Giannozzi}\ \emph {et~al.}(1991)\citenamefont {Giannozzi}, \citenamefont {De~Gironcoli}, \citenamefont {Pavone},\ and\ \citenamefont {Baroni}}]{giannozzi1991ab}%
  \BibitemOpen
  \bibfield  {author} {\bibinfo {author} {\bibfnamefont {P.}~\bibnamefont {Giannozzi}}, \bibinfo {author} {\bibfnamefont {S.}~\bibnamefont {De~Gironcoli}}, \bibinfo {author} {\bibfnamefont {P.}~\bibnamefont {Pavone}},\ and\ \bibinfo {author} {\bibfnamefont {S.}~\bibnamefont {Baroni}},\ }\href@noop {} {\bibfield  {journal} {\bibinfo  {journal} {Physical Review B}\ }\textbf {\bibinfo {volume} {43}},\ \bibinfo {pages} {7231} (\bibinfo {year} {1991})}\BibitemShut {NoStop}%
\bibitem [{\citenamefont {Chang}\ \emph {et~al.}(2010)\citenamefont {Chang}, \citenamefont {Chang},\ and\ \citenamefont {Chuang}}]{Changkp2010}%
  \BibitemOpen
  \bibfield  {author} {\bibinfo {author} {\bibfnamefont {G.-E.}\ \bibnamefont {Chang}}, \bibinfo {author} {\bibfnamefont {S.-W.}\ \bibnamefont {Chang}},\ and\ \bibinfo {author} {\bibfnamefont {S.~L.}\ \bibnamefont {Chuang}},\ }\href {https://doi.org/10.1109/JQE.2010.2059000} {\bibfield  {journal} {\bibinfo  {journal} {IEEE Journal of Quantum Electronics}\ }\textbf {\bibinfo {volume} {46}},\ \bibinfo {pages} {1813} (\bibinfo {year} {2010})}\BibitemShut {NoStop}%
\bibitem [{\citenamefont {Lu~Low}\ \emph {et~al.}(2012)\citenamefont {Lu~Low}, \citenamefont {Yang}, \citenamefont {Han}, \citenamefont {Fan},\ and\ \citenamefont {Yeo}}]{LuLowkp2012}%
  \BibitemOpen
  \bibfield  {author} {\bibinfo {author} {\bibfnamefont {K.}~\bibnamefont {Lu~Low}}, \bibinfo {author} {\bibfnamefont {Y.}~\bibnamefont {Yang}}, \bibinfo {author} {\bibfnamefont {G.}~\bibnamefont {Han}}, \bibinfo {author} {\bibfnamefont {W.}~\bibnamefont {Fan}},\ and\ \bibinfo {author} {\bibfnamefont {Y.-C.}\ \bibnamefont {Yeo}},\ }\href {https://doi.org/10.1063/1.4767381} {\bibfield  {journal} {\bibinfo  {journal} {Journal of Applied Physics}\ }\textbf {\bibinfo {volume} {112}},\ \bibinfo {pages} {103715} (\bibinfo {year} {2012})},\ \Eprint {https://arxiv.org/abs/https://doi.org/10.1063/1.4767381} {https://doi.org/10.1063/1.4767381} \BibitemShut {NoStop}%
\bibitem [{\citenamefont {Polak}\ \emph {et~al.}(2017)\citenamefont {Polak}, \citenamefont {Scharoch},\ and\ \citenamefont {Kudrawiec}}]{Polak_2017}%
  \BibitemOpen
  \bibfield  {author} {\bibinfo {author} {\bibfnamefont {M.~P.}\ \bibnamefont {Polak}}, \bibinfo {author} {\bibfnamefont {P.}~\bibnamefont {Scharoch}},\ and\ \bibinfo {author} {\bibfnamefont {R.}~\bibnamefont {Kudrawiec}},\ }\href {https://doi.org/10.1088/1361-6463/aa67bf} {\bibfield  {journal} {\bibinfo  {journal} {Journal of Physics D: Applied Physics}\ }\textbf {\bibinfo {volume} {50}},\ \bibinfo {pages} {195103} (\bibinfo {year} {2017})}\BibitemShut {NoStop}%
\bibitem [{\citenamefont {Rainko}\ \emph {et~al.}(2018)\citenamefont {Rainko}, \citenamefont {Ikonic}, \citenamefont {Vukmirovi{\'c}}, \citenamefont {Stange}, \citenamefont {von~den Driesch}, \citenamefont {Gr{\"u}tzmacher},\ and\ \citenamefont {Buca}}]{rainko2018investigation}%
  \BibitemOpen
  \bibfield  {author} {\bibinfo {author} {\bibfnamefont {D.}~\bibnamefont {Rainko}}, \bibinfo {author} {\bibfnamefont {Z.}~\bibnamefont {Ikonic}}, \bibinfo {author} {\bibfnamefont {N.}~\bibnamefont {Vukmirovi{\'c}}}, \bibinfo {author} {\bibfnamefont {D.}~\bibnamefont {Stange}}, \bibinfo {author} {\bibfnamefont {N.}~\bibnamefont {von~den Driesch}}, \bibinfo {author} {\bibfnamefont {D.}~\bibnamefont {Gr{\"u}tzmacher}},\ and\ \bibinfo {author} {\bibfnamefont {D.}~\bibnamefont {Buca}},\ }\href@noop {} {\bibfield  {journal} {\bibinfo  {journal} {Scientific reports}\ }\textbf {\bibinfo {volume} {8}},\ \bibinfo {pages} {1} (\bibinfo {year} {2018})}\BibitemShut {NoStop}%
\bibitem [{\citenamefont {Eißfeller}(2012)}]{Eisfeller2012}%
  \BibitemOpen
  \bibfield  {author} {\bibinfo {author} {\bibfnamefont {T.}~\bibnamefont {Eißfeller}},\ }\href@noop {} {\emph {\bibinfo {title} {Theory of the Electronic Structure of {{Quantum}} Dots in External Fields}}},\ \bibinfo {edition} {1st}\ ed.,\ \bibinfo {series} {Selected {{Topics}} of {{Semiconductor Physics}} and {{Technology}}}\ No.\ \bibinfo {number} {Vol. 146}\ (\bibinfo  {publisher} {{Verein zur Förderung des Walter-Schottky-Inst. der Techn. Univ. München}},\ \bibinfo {year} {2012})\BibitemShut {NoStop}%
\bibitem [{\citenamefont {Yamamoto}\ \emph {et~al.}(2018)\citenamefont {Yamamoto}, \citenamefont {Ueno},\ and\ \citenamefont {Tsukagoshi}}]{yamamoto2018pronounced}%
  \BibitemOpen
  \bibfield  {author} {\bibinfo {author} {\bibfnamefont {M.}~\bibnamefont {Yamamoto}}, \bibinfo {author} {\bibfnamefont {K.}~\bibnamefont {Ueno}},\ and\ \bibinfo {author} {\bibfnamefont {K.}~\bibnamefont {Tsukagoshi}},\ }\href@noop {} {\bibfield  {journal} {\bibinfo  {journal} {Applied Physics Letters}\ }\textbf {\bibinfo {volume} {112}},\ \bibinfo {pages} {181902} (\bibinfo {year} {2018})}\BibitemShut {NoStop}%
\bibitem [{\citenamefont {Fang}\ and\ \citenamefont {Hu}(2017)}]{fang2017photogating}%
  \BibitemOpen
  \bibfield  {author} {\bibinfo {author} {\bibfnamefont {H.}~\bibnamefont {Fang}}\ and\ \bibinfo {author} {\bibfnamefont {W.}~\bibnamefont {Hu}},\ }\href@noop {} {\bibfield  {journal} {\bibinfo  {journal} {Advanced science}\ }\textbf {\bibinfo {volume} {4}},\ \bibinfo {pages} {1700323} (\bibinfo {year} {2017})}\BibitemShut {NoStop}%
\bibitem [{\citenamefont {Jie}\ \emph {et~al.}(2006)\citenamefont {Jie}, \citenamefont {Zhang}, \citenamefont {Jiang}, \citenamefont {Meng}, \citenamefont {Li},\ and\ \citenamefont {Lee}}]{jie2006photoconductive}%
  \BibitemOpen
  \bibfield  {author} {\bibinfo {author} {\bibfnamefont {J.}~\bibnamefont {Jie}}, \bibinfo {author} {\bibfnamefont {W.}~\bibnamefont {Zhang}}, \bibinfo {author} {\bibfnamefont {Y.}~\bibnamefont {Jiang}}, \bibinfo {author} {\bibfnamefont {X.}~\bibnamefont {Meng}}, \bibinfo {author} {\bibfnamefont {Y.}~\bibnamefont {Li}},\ and\ \bibinfo {author} {\bibfnamefont {S.}~\bibnamefont {Lee}},\ }\href@noop {} {\bibfield  {journal} {\bibinfo  {journal} {Nano letters}\ }\textbf {\bibinfo {volume} {6}},\ \bibinfo {pages} {1887} (\bibinfo {year} {2006})}\BibitemShut {NoStop}%
\bibitem [{\citenamefont {Zheng}\ \emph {et~al.}(2018)\citenamefont {Zheng}, \citenamefont {Fang}, \citenamefont {Long}, \citenamefont {Wu}, \citenamefont {Wang}, \citenamefont {Gong}, \citenamefont {Wu}, \citenamefont {Ho}, \citenamefont {Liao},\ and\ \citenamefont {Hu}}]{zheng2018high}%
  \BibitemOpen
  \bibfield  {author} {\bibinfo {author} {\bibfnamefont {D.}~\bibnamefont {Zheng}}, \bibinfo {author} {\bibfnamefont {H.}~\bibnamefont {Fang}}, \bibinfo {author} {\bibfnamefont {M.}~\bibnamefont {Long}}, \bibinfo {author} {\bibfnamefont {F.}~\bibnamefont {Wu}}, \bibinfo {author} {\bibfnamefont {P.}~\bibnamefont {Wang}}, \bibinfo {author} {\bibfnamefont {F.}~\bibnamefont {Gong}}, \bibinfo {author} {\bibfnamefont {X.}~\bibnamefont {Wu}}, \bibinfo {author} {\bibfnamefont {J.~C.}\ \bibnamefont {Ho}}, \bibinfo {author} {\bibfnamefont {L.}~\bibnamefont {Liao}},\ and\ \bibinfo {author} {\bibfnamefont {W.}~\bibnamefont {Hu}},\ }\href@noop {} {\bibfield  {journal} {\bibinfo  {journal} {ACS nano}\ }\textbf {\bibinfo {volume} {12}},\ \bibinfo {pages} {7239} (\bibinfo {year} {2018})}\BibitemShut {NoStop}%
\bibitem [{\citenamefont {Wu}\ \emph {et~al.}(2011)\citenamefont {Wu}, \citenamefont {Dai}, \citenamefont {Ye}, \citenamefont {Yin},\ and\ \citenamefont {Dai}}]{wu2011fast}%
  \BibitemOpen
  \bibfield  {author} {\bibinfo {author} {\bibfnamefont {P.}~\bibnamefont {Wu}}, \bibinfo {author} {\bibfnamefont {Y.}~\bibnamefont {Dai}}, \bibinfo {author} {\bibfnamefont {Y.}~\bibnamefont {Ye}}, \bibinfo {author} {\bibfnamefont {Y.}~\bibnamefont {Yin}},\ and\ \bibinfo {author} {\bibfnamefont {L.}~\bibnamefont {Dai}},\ }\href@noop {} {\bibfield  {journal} {\bibinfo  {journal} {Journal of Materials Chemistry}\ }\textbf {\bibinfo {volume} {21}},\ \bibinfo {pages} {2563} (\bibinfo {year} {2011})}\BibitemShut {NoStop}%
\bibitem [{\citenamefont {Koelling}\ \emph {et~al.}(2017)\citenamefont {Koelling}, \citenamefont {Li}, \citenamefont {Cavalli}, \citenamefont {Assali}, \citenamefont {Car}, \citenamefont {Gazibegovic}, \citenamefont {Bakkers},\ and\ \citenamefont {Koenraad}}]{koelling2017atom}%
  \BibitemOpen
  \bibfield  {author} {\bibinfo {author} {\bibfnamefont {S.}~\bibnamefont {Koelling}}, \bibinfo {author} {\bibfnamefont {A.}~\bibnamefont {Li}}, \bibinfo {author} {\bibfnamefont {A.}~\bibnamefont {Cavalli}}, \bibinfo {author} {\bibfnamefont {S.}~\bibnamefont {Assali}}, \bibinfo {author} {\bibfnamefont {D.}~\bibnamefont {Car}}, \bibinfo {author} {\bibfnamefont {S.}~\bibnamefont {Gazibegovic}}, \bibinfo {author} {\bibfnamefont {E.}~\bibnamefont {Bakkers}},\ and\ \bibinfo {author} {\bibfnamefont {P.}~\bibnamefont {Koenraad}},\ }\href@noop {} {\bibfield  {journal} {\bibinfo  {journal} {Nano letters}\ }\textbf {\bibinfo {volume} {17}},\ \bibinfo {pages} {599} (\bibinfo {year} {2017})}\BibitemShut {NoStop}%
\end{thebibliography}%


\begin{thebibliography}{3}%
\makeatletter
\providecommand \@ifxundefined [1]{%
 \@ifx{#1\undefined}
}%
\providecommand \@ifnum [1]{%
 \ifnum #1\expandafter \@firstoftwo
 \else \expandafter \@secondoftwo
 \fi
}%
\providecommand \@ifx [1]{%
 \ifx #1\expandafter \@firstoftwo
 \else \expandafter \@secondoftwo
 \fi
}%
\providecommand \natexlab [1]{#1}%
\providecommand \enquote  [1]{``#1''}%
\providecommand \bibnamefont  [1]{#1}%
\providecommand \bibfnamefont [1]{#1}%
\providecommand \citenamefont [1]{#1}%
\providecommand \href@noop [0]{\@secondoftwo}%
\providecommand \href [0]{\begingroup \@sanitize@url \@href}%
\providecommand \@href[1]{\@@startlink{#1}\@@href}%
\providecommand \@@href[1]{\endgroup#1\@@endlink}%
\providecommand \@sanitize@url [0]{\catcode `\\12\catcode `\$12\catcode `\&12\catcode `\#12\catcode `\^12\catcode `\_12\catcode `\%12\relax}%
\providecommand \@@startlink[1]{}%
\providecommand \@@endlink[0]{}%
\providecommand \url  [0]{\begingroup\@sanitize@url \@url }%
\providecommand \@url [1]{\endgroup\@href {#1}{\urlprefix }}%
\providecommand \urlprefix  [0]{URL }%
\providecommand \Eprint [0]{\href }%
\providecommand \doibase [0]{https://doi.org/}%
\providecommand \selectlanguage [0]{\@gobble}%
\providecommand \bibinfo  [0]{\@secondoftwo}%
\providecommand \bibfield  [0]{\@secondoftwo}%
\providecommand \translation [1]{[#1]}%
\providecommand \BibitemOpen [0]{}%
\providecommand \bibitemStop [0]{}%
\providecommand \bibitemNoStop [0]{.\EOS\space}%
\providecommand \EOS [0]{\spacefactor3000\relax}%
\providecommand \BibitemShut  [1]{\csname bibitem#1\endcsname}%
\let\auto@bib@innerbib\@empty
\bibitem [{\citenamefont {Bouthillier}\ \emph {et~al.}(2020)\citenamefont {Bouthillier}, \citenamefont {Assali}, \citenamefont {Nicolas},\ and\ \citenamefont {Moutanabbir}}]{bouthillier2020decoupling}%
  \BibitemOpen
  \bibfield  {author} {\bibinfo {author} {\bibfnamefont {{\'E}.}~\bibnamefont {Bouthillier}}, \bibinfo {author} {\bibfnamefont {S.}~\bibnamefont {Assali}}, \bibinfo {author} {\bibfnamefont {J.}~\bibnamefont {Nicolas}},\ and\ \bibinfo {author} {\bibfnamefont {O.}~\bibnamefont {Moutanabbir}},\ }\href@noop {} {\bibfield  {journal} {\bibinfo  {journal} {Semiconductor Science and Technology}\ }\textbf {\bibinfo {volume} {35}},\ \bibinfo {pages} {095006} (\bibinfo {year} {2020})}\BibitemShut {NoStop}%
\bibitem [{\citenamefont {Nelin}\ and\ \citenamefont {Nilsson}(1972)}]{nelin1972phonon}%
  \BibitemOpen
  \bibfield  {author} {\bibinfo {author} {\bibfnamefont {G.}~\bibnamefont {Nelin}}\ and\ \bibinfo {author} {\bibfnamefont {G.}~\bibnamefont {Nilsson}},\ }\href@noop {} {\bibfield  {journal} {\bibinfo  {journal} {Physical Review B}\ }\textbf {\bibinfo {volume} {5}},\ \bibinfo {pages} {3151} (\bibinfo {year} {1972})}\BibitemShut {NoStop}%
\bibitem [{\citenamefont {Giannozzi}\ \emph {et~al.}(1991)\citenamefont {Giannozzi}, \citenamefont {De~Gironcoli}, \citenamefont {Pavone},\ and\ \citenamefont {Baroni}}]{giannozzi1991ab}%
  \BibitemOpen
  \bibfield  {author} {\bibinfo {author} {\bibfnamefont {P.}~\bibnamefont {Giannozzi}}, \bibinfo {author} {\bibfnamefont {S.}~\bibnamefont {De~Gironcoli}}, \bibinfo {author} {\bibfnamefont {P.}~\bibnamefont {Pavone}},\ and\ \bibinfo {author} {\bibfnamefont {S.}~\bibnamefont {Baroni}},\ }\href@noop {} {\bibfield  {journal} {\bibinfo  {journal} {Physical Review B}\ }\textbf {\bibinfo {volume} {43}},\ \bibinfo {pages} {7231} (\bibinfo {year} {1991})}\BibitemShut {NoStop}%
\end{thebibliography}%
\bibliographystyle{apsrev4-2} 


\end{document}


\def\figureautorefname{Fig.} 
\def\tableautorefname{Table} 
\newcommand{\fref}[2]{\autoref{#1}\textcolor{blue}{#2}}

\newcommand{\paperSection}[2][normal]{
    \ifthenelse{\equal{#1}{normal}}{
        \medskip\noindent{\textbf{#2}}\newline
    }{
        \noindent\textbf{#2}\newline
    }
}

 
\newcommand{\RomanNumeralCaps}[1]
    {\MakeUppercase{\romannumeral #1}}

\newcommand{\polydept}{Department of Engineering Physics, \'Ecole Polytechnique de Montr\'eal, C.P. 6079, Succ. Centre-Ville, Montr\'eal, Qu\'ebec, Canada H3C 3A7}

\newcommand{\citechange}[1]{\textcolor{red}{#1}}
\newcommand{\SiGeSn}[3]{Si$_{#1}$Ge$_{#2}$Sn$_{#3}$}
\newcommand{\SipGeSn}[3]{(Si)$_{#1}$Ge$_{#2}$Sn$_{#3}$}
\newcommand{\SiGe}[2]{Si$_{#1}$Ge$_{#2}$}
\newcommand{\quot}[1]{``#1''} 
\newcommand{\ac}[1]{\gls*{#1}}
\newcommand{\acc}[1]{\Gls*{#1}}
\newcommand{\RNum}[1]{\uppercase\expandafter{\romannumeral #1\relax}}

\newcommand{\citesupp}[1]{
    \footnote{See Supplemental Material at \textbf{[URL will be inserted by publisher]} for \textcolor{red}{details of the experimental setup}, which includes Refs. #1}
}


\newcommand{\figCiteChange}[1]{\colorbox{yellow!50!white}{#1}}
\DeclareSIUnit{\million}{\text{million}}


\newcommand\vertarrowbox[3][3ex]{%
  \begin{array}[t]{@{}c@{}} #2 \vspace{1ex}\\
  \left\uparrow\vcenter{\hrule height #1}\right.\kern-\nulldelimiterspace\\
  \makebox[0pt]{#3}
  \end{array}%
}

\NewEnviron{myequation}[1]{%
\begin{equation}
\scalebox{#1}{$\BODY$}
\end{equation}
}

\newcommand{\suppfref}[2]{\autoref{#1}\textcolor{blue}{#2}}

\newcommand\suppFigRef[3]{%
    \ifthenelse{\equal{\includeSI}{1}}{%
        \let\tempfigureautorefname\figureautorefname%
        \renewcommand\figureautorefname{Fig.}%
        \suppfref{#1}{#2}%
        \let\figureautorefname\tempfigureautorefname%
    }{\textcolor{blue}{Fig. S#3}}%
}

\newcommand\suppFigsRef[3]{%
    \ifthenelse{\equal{\includeSI}{1}}{%
        \let\tempfigureautorefname\figureautorefname%
        \renewcommand\figureautorefname{Figs.}%
        \suppfref{#1}{#2}%
        \let\figureautorefname\tempfigureautorefname%
    }{\textcolor{blue}{Figs. S#3}}%
}

\newcommand\suppTabRef[3]{%
    \ifthenelse{\equal{\includeSI}{1}}{%
        \let\temptableautorefname\tableautorefname%
        \renewcommand\tableautorefname{Table}%
        \suppfref{#1}{#2}%
        \let\tableautorefname\temptableautorefname%
    }{\textcolor{blue}{Table S#3}}%
}

\def\blankpage{%
      \clearpage%
      \thispagestyle{empty}%
      \null%
      \clearpage}

\epstopdfsetup{update} 
\epstopdfsetup{outdir=./figures/}
\epstopdfsetup{suffix=-generated}

\newacronym{se}{SE}{spectroscopic ellipsometry}
\newacronym{apt}{APT}{atom probe tomography}
\newacronym{fib}{FIB}{focused ion beam}
\newacronym{hrxrd}{HRXRD}{high\hyp{}resolution xray diffraction}
\newacronym{rsm-xrd}{RSM\hyp{}XRD}{reciprocal space mapping X\hyp{}ray diffraction spectroscopy}
\newacronym{rsm}{RSM}{reciprocal space mapping}
\newacronym{fwhm}{FWHM}{full\hyp{}width half maximum}
\newacronym{edsx}{EDSX}{energy\hyp{}dispersive X\hyp{}ray spectroscopy}
\newacronym{tem}{TEM}{transmission electron microscope}
\newacronym{hrtem}{HRTEM}{High\hyp{}resolution transmission electron microscopy}
\newacronym{stem}{STEM}{scanning transmission electron microscopy}
\newacronym{mocvd}{MOCVD}{metal\hyp{}organic chemical vapor deposition}
\newacronym{cvd}{CVD}{chemical vapor deposition}
\newacronym{cmos}{CMOS}{complementary metal\hyp{}oxide\hyp{}semiconductor}
\newacronym{haadf-stem}{HAADF\hyp{}STEM}{high\hyp{}angle annular dark field scanning transmission electron microscopy}
\newacronym{rms}{RMS}{surface roughness}
\newacronym{eels}{EELS}{electron energy loss spectroscopy}
\newacronym{rta}{RTA}{rapid thermal annealing}
\newacronym{lta}{LTA}{laser thermal annealing}
\newacronym{ia}{IA}{isothermal annealing}
\newacronym{rheed}{RHEED}{reflection high-energy electron diffraction}
\newacronym{peem}{PEEM}{photoemission electron microscopy}
\newacronym{leed}{LEED}{low\hyp{}energy electron diffraction}
\newacronym{sims}{SIMS}{secondary ion mass spectroscopy}

\newacronym{afm}{AFM}{atomic force microscopy}
\newacronym[\glslongpluralkey={misfit dislocations}]{md}{MD}{misfit dislocation}
\newacronym{td}{TD}{threading dislocation}
\newacronym{fib}{FIB}{focused\hyp{}ion beam}
\newacronym{ctlm}{CTLM}{circular transfer length method}
\newacronym{tlm}{TLM}{transfer length method}
\newacronym{rtse}{RTSE}{room temperature spectroscopic ellipsometry}
\newacronym{vs}{VS}{virtual substrate}
\newacronym{pecvd}{PECVD}{plasma enhanced chemical vapor deposition}
\newacronym{aoi}{AOI}{angle of incidence}
\newacronym[\glslongpluralkey={critical\hyp{}points}]{cp}{CP}{critical\hyp{}point}
\newacronym{cppb}{CPPB}{critical point parabolic band}

\title{Mid-Infrared Detectors and Imagers Integrating All-Group IV Nanowires}

\author{Lu Luo} 
\affiliation{\polydept{}}

\author{Mahmoud RM Atalla}  
\affiliation{\polydept{}}

\author{Simone Assali} 
\affiliation{\polydept{}}

\author{Sebastian Koelling} 
\affiliation{\polydept{}}

\author{G\'erard Daligou} 
\affiliation{\polydept{}}

\author{Oussama Moutanabbir}
\affiliation{\polydept{}}

\maketitle
\tableofcontents
\newpage
\section{APT measurement of the nanowire with a thin and a thick Ge core}

\begin{figure}[htb]
    \centering
    \includegraphics[width=.6\textwidth]{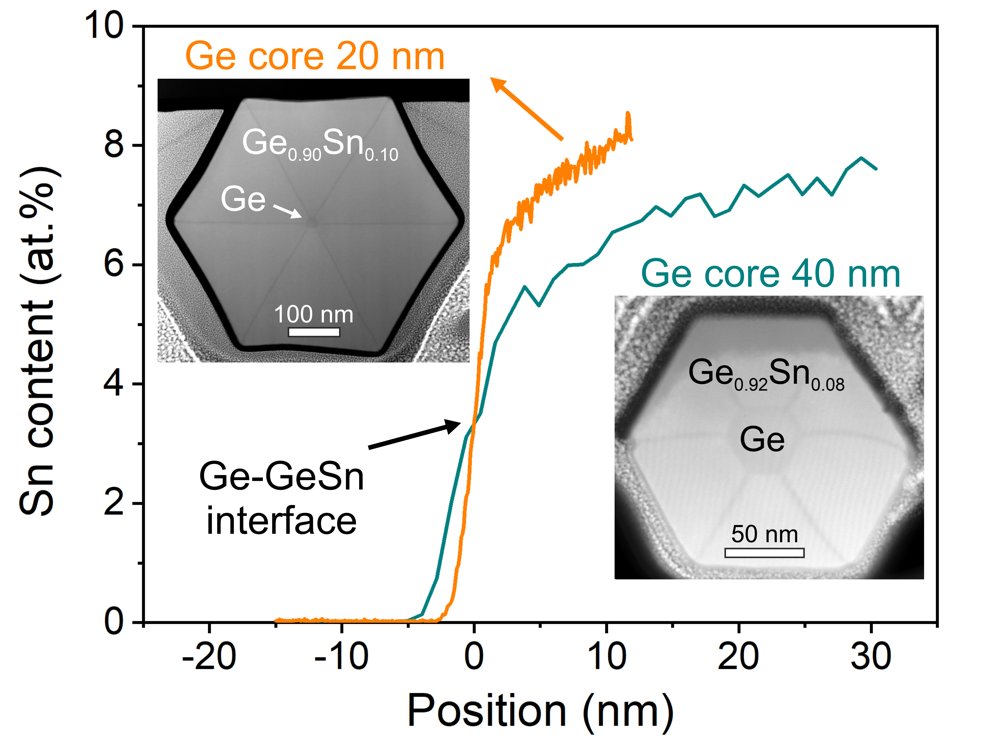}
    \caption{Atom probe tomography (APT) showing Sn radial profiles for Ge/GeSn core/shell nanowires grown using thin (orange) and thick (dark cyan) Ge core. Inset: Representative cross-sectional transmission electron microscope (TEM) images of these two nanowires. Note that the growth on thin Ge core is associated with a faster increase in Sn content.}
    \label{fig:figS1}
\end{figure}
Figure S1 compares the Sn radial profiles measured by atom probe tomography content close to Ge/GeSn interface for a nanowire with a thin core and another with a thick core. We note that for a thin core the Sn content increases rapidly as the GeSn shell growth progresses leading to higher and uniform radial composition. 

\newpage
\section{Raman spectrum of single Ge/Ge$_{0.89}$Sn$_{0.11}$ core/shell NW}

\begin{figure}[htb]
    \centering
    \includegraphics[width=.9\textwidth]{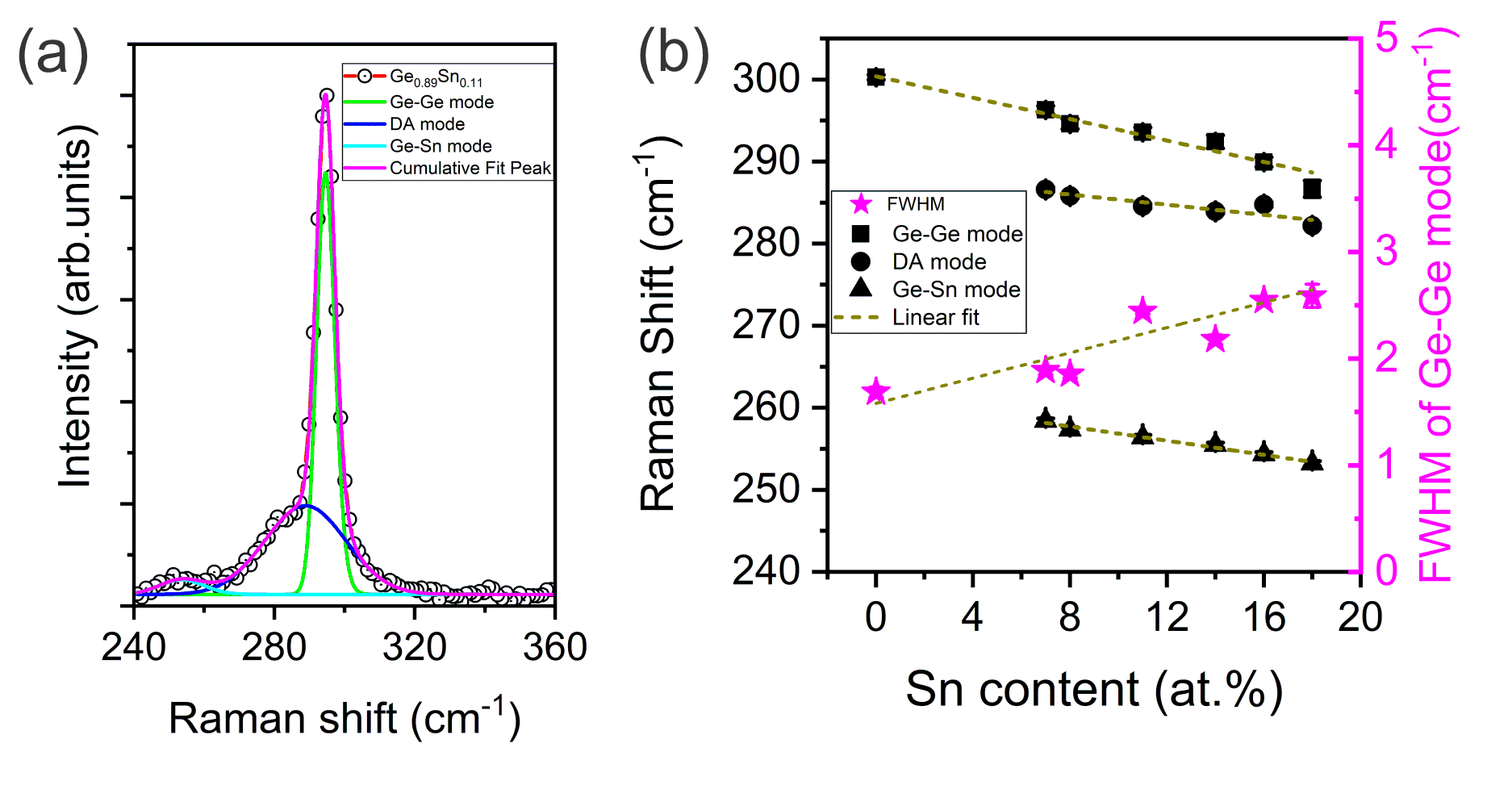}
    \caption{(a) Fit of the Ge-Ge, DA, and Ge-Sn modes for single Ge/Ge$_{0.89}$Sn$_{0.11}$ core/shell NW. (b) Ge-Ge, DA, Ge-Sn modes and the full width at half maximum (FWHM) of the Ge-Ge mode as the function of the Sn content.}
    \label{fig:figS2}
\end{figure}
To accurately identify the multiple modes, we fitted the spectrum with exponentially modified gaussian (EMG) functions. Fig.S2 (a) displays fitted Raman spectrum of the single Ge/Ge$_{0.89}$Sn$_{0.11}$ core/shell NW in the range of 240 cm$^{-1}$-360 cm$^{-1}$. In the Ge/Ge$_{1-x}$Sn$_{x}$ core/shell NWs, apart from the Ge-Ge mode located near 300 cm$^{-1}$ and Ge-Sn mode located around 250-260 cm$^{-1}$, an additional mode located in the range of 280-290 cm$^{-1}$ was detected as well. The additional mode, which corresponds to disorder-activated (DA) Raman scattering, has been also detected in GeSn thin films and GeSn NWs.\cite{bouthillier2020decoupling,nelin1972phonon,giannozzi1991ab} The intensity of the DA mode can change with the polarization direction of both the polarizer and the analyzer. Ge-Ge, Ge-Sn and the DA modes are all linearly shifted to a lower wavenumber with increasing Sn composition, while the full width at half maximum (FWHM) of Ge-Ge mode are slightly increased, which are summarized in Fig.S2 (b). The Sn-Sn modes in the individual Ge/Ge$_{1-x}$Sn$_{x}$ core/shell NWs are not clear because of the small volume of the material.
\newpage
\section{Photoresponse under 2330 nm laser}
\begin{figure}[htb]
    \centering
    \includegraphics[width=.9\textwidth]{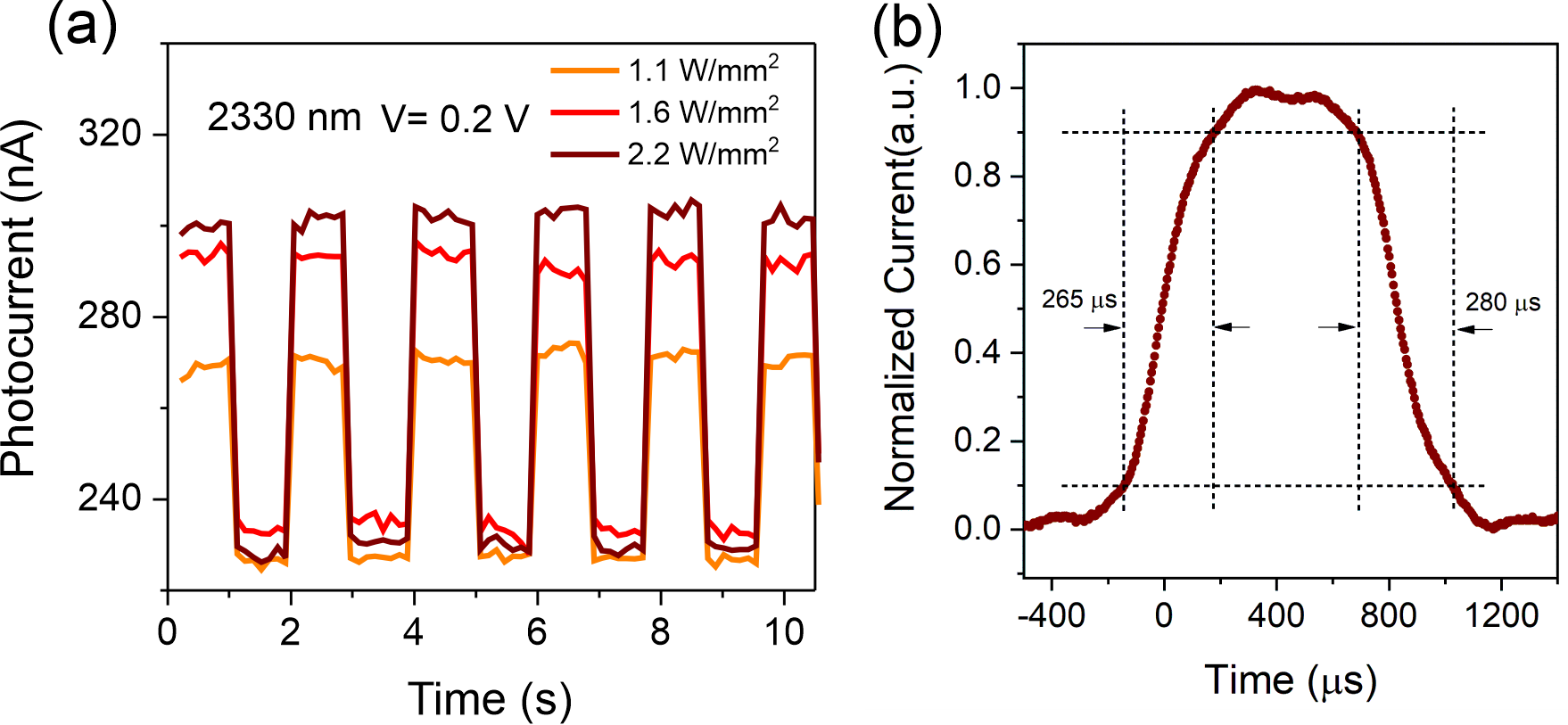}
    \caption{(a) Fit of the Ge-Ge, DA, and Ge-Sn modes for single Ge/Ge$_{0.89}$Sn$_{0.11}$ core/shell NW. (b) Ge-Ge, DA, Ge-Sn modes and the full width at half maximum (FWHM) of the Ge-Ge mode as the function of the Sn content.}
    \label{fig:figS3}
\end{figure}
Fig. S3(a) presents the transient photocurrent of the multiple NW device under the illumination of 2330 nm laser with different power densities. The rise time and the fall time at 2330 nm are 265 $\mu$s and 280 $\mu$s, respectively, as shown in Fig.S3 (b).

\newpage
\section{Photocurrent at different temperatures}
\begin{figure}[htb]
    \centering
    \includegraphics[width=.5\textwidth]{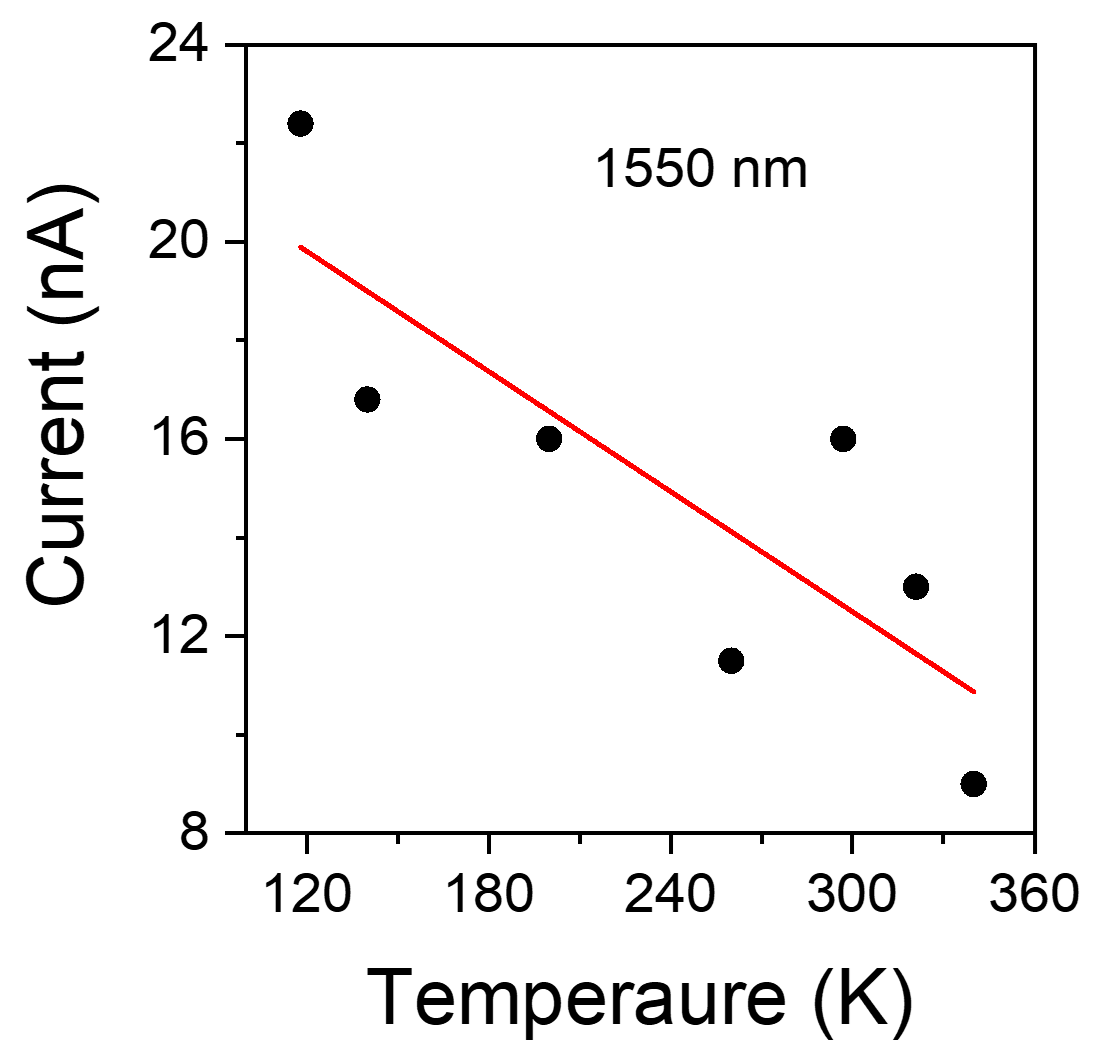}
    \caption{Photocurrent of the multiple NW device measured at different temperatures under 1550 nm illumination. }
    \label{fig:figS4}
\end{figure}
The obtained photocurrent slightly increases from 9 nA to 22 nA at 0.2 V bias as the temperature decreases from 340 K to 118 K.
\newpage
\section{Single-pixel imaging setup}
\begin{figure}[htb]
    \centering
    \includegraphics[width=.8\textwidth]{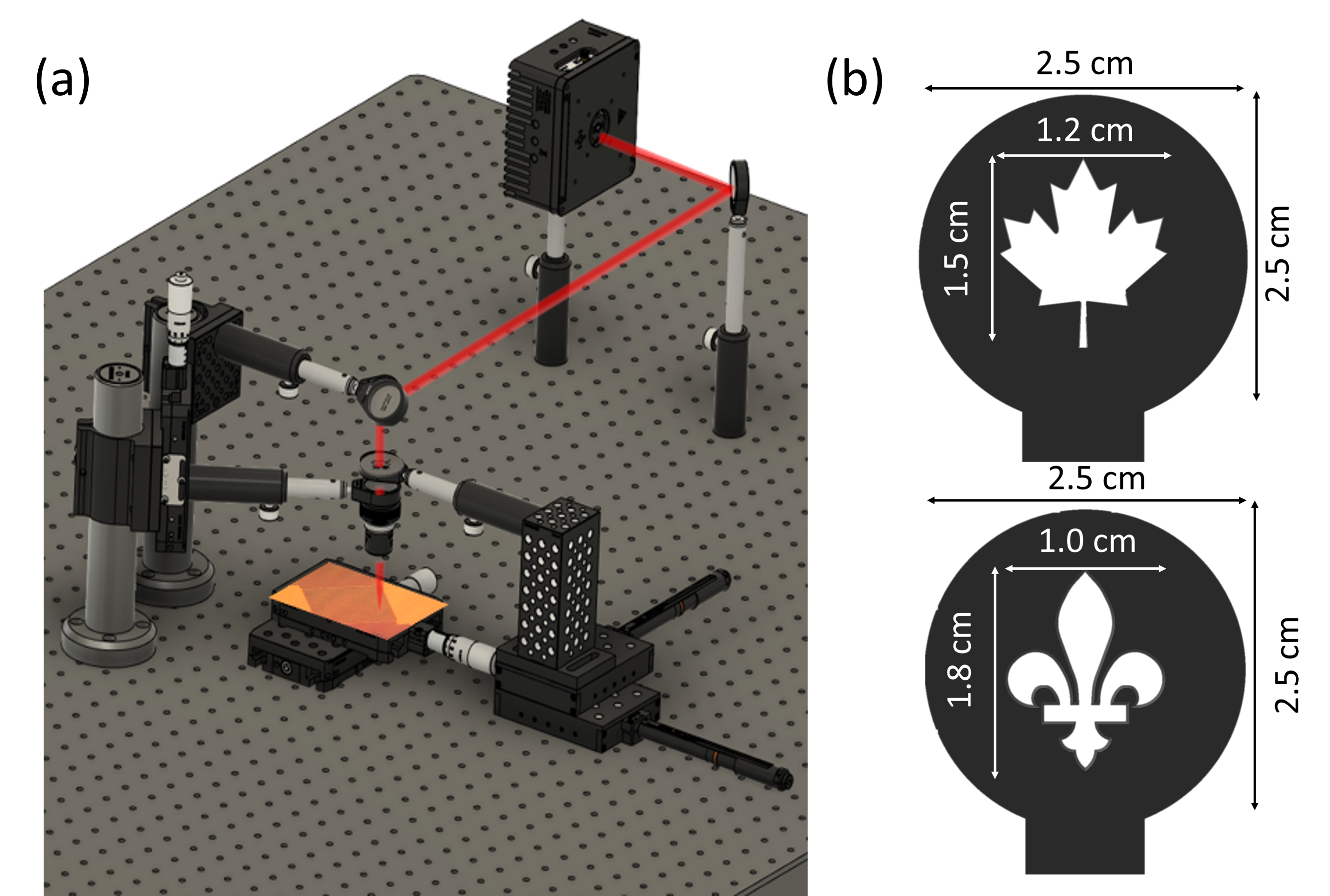}
    \caption{(a) Setup for NW single-pixel imaging measurement. (b) 3D-printed maple leaf and lily flower patterns.}
    \label{fig:figS5}
\end{figure}
The major components of the imaging setup include a light source, a focus lens, two reflective mirrors, and a motorized XY stage, as shown in Fig. S5(a). Fig. S5(b) are the 3D-printed maple leaf and lily flower patterns, where the area of the maple leaf is 1.2 cm × 1.5 cm and the area of the lily flower is 1.0 cm × 1.8 cm.


\bibliography{NW_SI.bib} 
\bibliographystyle{apsrev4-2} 

